\documentclass{aa}  
\usepackage{graphicx}
\usepackage{txfonts}
\begin{document}
    \title{VLA observations of water masers towards 6.7\,GHz methanol maser sources
}
   \author{A. Bartkiewicz
          \inst{1},
          M. Szymczak
          \inst{1},
          Y.M. Pihlstr\"om
          \inst{2,3},
          H.J. van Langevelde
          \inst{4,5},
          A. Brunthaler
          \inst{6},
          \and
          M.J. Reid\inst{7}
          }

   \institute{Toru\'n Centre for Astronomy, Nicolaus Copernicus
          University, Gagarina 11, 87-100 Toru\'n, Poland\\
          \email{annan@astro.uni.torun.pl}
\and      Department of Physics and Astronomy, MSC07 4220, University of New Mexico, Albuquerque, 
          NM 87131, USA
\and      National Radio Astronomy Observatory, 1003 Lopezville Road, Socorro, NM 87801, USA
\and      Joint Institute for VLBI in Europe, Postbus 2, 7990 AA Dwingeloo, The Netherlands
\and      Sterrewacht Leiden, Leiden University, Postbus 9513, 2300 RA Leiden, The Netherlands
\and      Max-Planck-Insitut f\"ur Radioastronomie, Auf dem H\"ugel 69, 53121 Bonn, Germany
\and      Harvard-Smithsonian Center for Astrophysics, 60 Garden Street, Cambridge, MA 02138, USA
          }

   \date{Received 2010 June 18; accepted 2010 September 09}

\authorrunning{Bartkiewicz et al.}
\titlerunning{VLA observations of water masers towards 6.7\,GHz methanol masers}

 
\abstract 
  {22\,GHz water and 6.7\,GHz methanol masers are usually thought as
  signposts of early stages of high-mass star formation but little is
  known about their associations and the physical environments they occur in.}  
  {To obtain accurate positions and morphologies of the water maser emission and
  relate them to the methanol maser emission recently mapped with 
  Very Long Baseline Interferometry.}  {A sample of 31 methanol maser sources 
  was searched for 22\,GHz water masers using the VLA and observed in the 6.7\,GHz
  methanol maser line with the 32\,m Torun dish simultaneously.}  {Water maser clusters 
  were detected towards 27 sites finding 15 new sources. The
  detection rate of water maser emission associated with methanol
  sources was as high as 71\%. In a large number of objects
  (18/21) the structure of water maser is well aligned with that of
  the extended emission at 4.5\,$\mu$m confirming the origin of water
  emission from outflows. The sources with methanol emission with ring-like
  morphologies, which likely trace a circumstellar disk/torus, either do not 
  show associated water masers or the distribution of water maser spots is 
  orthogonal to the major axis of the ring.}  
  {The two maser species are generally
  powered by the same high-mass young stellar object but probe
  different parts of its environment. The morphology of water and 
  methanol maser emission in a minority of
  sources is consistent with a scenario that 6.7\,GHz methanol masers
  trace a disc/torus around a protostar while the associated 22\,GHz
  water masers arise in outflows. The majority of sources in which  
  methanol maser emission is associated with the water maser 
  appears to trace outflows. The two types of associations might be related to different  
  evolutionary phases.}


   \maketitle
%

\section{Introduction}
Studies of high-mass star forming regions (HMSFRs) are difficult but
important in astrophysics as they are responsible for many
of the energetic phenomena we see in galaxies. However,
their large distances, heavy obscuration and rapidity of evolution make
observations challenging. Maser emission has become a unique 
tool to study massive star formation. Methanol masers at 6.7\,GHz as well as water 
masers at 22\,GHz have been recognized as tracers of massive star formation 
(e.g., Caswell et
al.\ \cite{caswell95}; Menten \cite{menten91}; Sridharan et
al.\ \cite{sridharan02}; Urquhart et al.\ \cite{urquhart10}). Moreover,
both maser species are found associated with the
very early stage of a protostar, when it still accretes and 
before it begins to ionise the surrounding medium. These masers are often 
detectable before an ultra--compact H\,{\small II} region is 
seen at cm wavelengths.

Studies of maser emission at milliarcsecond scale, using Very Long Baseline
Interferometry (VLBI) techniques, reveal a wide range of morphologies of 6.7\,GHz 
methanol masers. They can form simple structures (a single spot), lie in linear
structures or arcs, or are distributed randomly without any apparent regularity (e.g., 
Minier et al. \cite{minier00}; Norris et al. \cite{norris98}; Phillips et al. 
\cite{phillips98}; Walsh et al. \cite{walsh98}). 
However, it is still unclear where and how they are produced. {\it Are they
related with disks/tori around young massive protostars or found in outflows
or shocks?} (e.g., Dodson et al. \cite{dodson04}; 
Minier et al. \cite{minier00}; Walsh et al. \cite{walsh98}). 
Detailed studies of particular sources reveal further clues to 
the origin of methanol masers. Unfortunately, they are not always
consistent with one scenario. High angular 
resolution mid-infrared (MIR) observations by De Buizer \& Minier
(\cite{debuizer05m}) revealed that the outflow
scenario is more plausible in the case of NGC~7538\,IRS\,1, where
the linear structure of methanol masers had been suggested 
as originating in an edge-on Keplerian disc (Pestalozzi et al.
\cite{pestalozzi04}). On the other hand, van der Walt et al. (\cite{walt07}) 
argued that a simple Keplerian-like disc model was more consistent with the 
observed kinematics of methanol maser spots in linear structures than the 
shock model proposed by Dodson et al. (\cite{dodson04}). 

A relatively high detection rate of water masers toward methanol masers 
is confirmed with single-dish studies. Szymczak et al. (\cite{szymczak05}) 
observed 79 targets with 6.7\,GHz methanol maser emission and detected the 
22\,GHz water line in 52\% cases. Similarly, Sridharan et al. 
(\cite{sridharan02}) reported a detection rate of 42\% for the sample of 
69 HMSFRs. In interferometric investigations Beuther et al. (\cite{beuther02}) 
obtained a 62\% detection rate of water masers toward methanol masers. 
Recently, Breen et al. (\cite{breen10}) searched 379 water masers and found 
methanol emission in $\approx52$\% of the sources. 
Although different excitation conditions are required for both molecules, their 
origin is in some sense dependent and likely related to the same powering 
source. 

There are few HMSFRs with detailed studies of 
methanol--water maser associations. For example Pillai et al. (\cite{pillai06}) observed 
the HMSFR G11.11$-$0.12 over a wide wavelength range. They reported that 
methanol masers were associated with an accretion disc driving an outflow 
traced by water maser emission. Moscadelli et al. (\cite{moscadelli07}) 
explored HMSFR G24.78$+$0.08 and showed that water masers trace a fast 
expanding shell closely surrounding a hyper-compact H\,{\small
II} region. Methanol masers were proposed to have emerged in a rotating
toroid lying radially outward of the H\,{\small II} region. However, there is a 
lack of data for a large sample of methanol and water masers at high angular 
resolution with a few mJy sensitivity to get better statistics on the two
types of associations. 

We have recently completed a survey of 31 sources at 6.7\,GHz using the 
European VLBI Network (EVN) (Bartkiewicz et al. \cite{bartkiewicz09}). Due 
to the high angular and spectral resolution as well as the high sensitivity
we have discovered nine sources (29\% of the sample) 
with ring-like maser distributions (with a typical major axis of 
0\farcs19). These ring-like structures
strongly suggest the existence of a central object, and could provide
a clue to its nature. Each source with ring-like morphology
coincides within 1\arcsec\,\, with a MIR object (from the GLIMPSE survey) that 
has an excess of 4.5\,$\mu$m emission, which is evidence 
for shocked regions (e.g., Cyganowski et al. 2008). This suggests 
that even ring-like structures can arise due to shock waves or in outflows. 
In order to answer the question {\it what are these structures?}, we 
initiated wide and detailed studies of that sample of methanol maser
sources. Here, we present the first results of our investigation of the
presence, position, and distribution of water maser emission toward 
6.7\,GHz methanol maser emission. We used the NRAO Very Large Array (VLA) 
to search for water masers near the locations of 6.7~GHz methanol masers
and, if detected, to compare the positions of the two masing species.

\section{Observations and data reduction}
\subsection{VLA observations}
To investigate the relationship between water and methanol
masers in HMSFRs, our sample of 31 methanol maser sources
(Table~\ref{table:1}) was observed at 22.23508\,GHz using the VLA in
CnB configuration in two 12\,h runs on 2009 June 4 and 5 (the project
AB1324). A spectral line mode with a single IF and 6.25\,MHz bandwidth divided
into 128 spectral channels was used, yielding a velocity coverage of
84\,km\,s$^{-1}$ and a channel spacing of 0.65\,km\,s$^{-1}$. The
pointing positions were defined as the coordinates of the brightest
6.7\,GHz methanol maser component (Table~\ref{table:1}) and the
bandpass was centred at the methanol maser peak velocity
taken from Bartkiewicz et al.\ (\cite{bartkiewicz09}, their
Table 5). 3C\,286 was used as the primary flux density calibrator for
all targets. We used two secondary calibrators (J1851$+$0035 and
J1832$-$1035) to monitor changes in interferometer amplitude and phase;
these were selected from the VLA calibrator catalog to be near the targets
(Table~\ref{table:1}). We allocated 50~s for observation of the secondary
calibrator, followed by 250\,s for the maser source. These times included 
slew and on-source integration times. In total each target was observed for 35\,min,
resulting in about 29\,min on-source integration time.

The data reduction was carried out following the standard recipes
recommended in Appendix B of the AIPS cookbook\footnote{See
  http://www.nrao.edu/aips/coobook.html}. The amplitude and phase
errors of 3C\,286 were corrected using the default source model and
3C\,286 was subsequently used to derive the secondary-calibrator flux
densities. The antenna gains were calibrated using the
secondary-calibrator data. A few bad data points were flagged and images 
(512$\times$512 pixels with pixel size of 0\farcs15) were
created using natural weighting. The noise levels in the maps
and the synthesized beams are listed in
Table~\ref{table:1}. The analysis of maser properties was carried out
using maps centred on the position of the brightest water maser
spots. 

We estimate that, with the relatively stable weather conditions during our
observations, position errors of water maser spots are dominated by the
errors of the secondary-calibrator positions, which could be as large as
0\farcs15 for these two calibrators. However, the relative position
uncertainties are much better ($\approx$10\,mas).

\subsection{32\,m dish observations}
The same sample was observed in the 6.7\,GHz methanol maser line using
the Torun 32\,m telescope over 20 days in June 2009 nearly simultaneously 
with the VLA H$_2$O observations. The telescope characteristics and calibration
procedures were described in Szymczak et al.\ (\cite{szymczak02}). The
spectra were taken in frequency switching mode with a resulting
spectral channel spacing of 0.04\,km\,s$^{-1}$ and sensitivity of
$\sim$0.6\,Jy (3$\sigma$). The accuracy of the absolute flux density calibration was
better than 15\%.

\begin{table*}
\caption{6.7\,GHz methanol maser sites searched for the 22\,GHz water maser emission}
\label{table:1}     
\centering          
\begin{tabular}{lllcccc}
  \hline\hline
  Source$^*$& \multicolumn{2}{c}{Position of 6.7\,GHz masers} & V$_{\rm p}$&
  Secondary calibrator & Synthesized beam & Rms noise \\
  Gll.lll$\pm$bb.bbb& RA(h m s) & Dec(\degr \,\arcmin \, \arcsec) & (km\,s$^{-1}$)&
  calibrator & maj, min;  PA (\arcsec, \arcsec;$^{\rm o}$) & per channel (mJy\,beam$^{-1}$)\\
  \hline                    
  G21.407$-$00.254 & 18 31 06.33794 & $-$10 21 37.4108 & 89.0&J1832$-$1035 & 1.19, 0.49;$+$61 & 2\\
  G22.335$-$00.155 & 18 32 29.40704 & $-$09 29 29.6840 & 35.6&J1832$-$1035 & 1.07, 0.72;$+$48 & 4\\
  G22.357$+$00.066 & 18 31 44.12055 & $-$09 22 12.3129 & 79.7&J1832$-$1035 & 1.36, 0.73;$+$39 & 3\\
  G23.207$-$00.377 & 18 34 55.21212 & $-$08 49 14.8926 & 77.1&J1832$-$1035 & 1.05, 0.85;$+$2 & 4\\
  G23.389$+$00.185 & 18 33 14.32477 & $-$08 23 57.4723 & 75.4&J1832$-$1035 & 0.88, 0.84;$+$83 & 3\\
  G23.657$-$00.127 & 18 34 51.56482 & $-$08 18 21.3045 & 82.6&J1832$-$1035 & 1.03, 0.55;$-$77.5 & 3\\
  G23.707$-$00.198 & 18 35 12.36600 & $-$08 17 39.3577 & 79.2&J1832$-$1035 & 1.22, 0.84;$-$37 & 4\\
  G23.966$-$00.109 & 18 35 22.21469 & $-$08 01 22.4698 & 70.9&J1832$-$1035 & 1.70, 0.72;$-$44 & 5\\
  G24.148$-$00.009 & 18 35 20.94266 & $-$07 48 55.6745 & 17.8&J1832$-$1035 & 1.65, 0.52;$-$60 & 4\\
  G24.541$+$00.312 & 18 34 55.72152 & $-$07 19 06.6504 &105.7&J1832$-$1035 & 1.89, 0.74;$-$48 & 6\\
  G24.634$-$00.324 & 18 37 22.71271 & $-$07 31 42.1439 & 35.4&J1832$-$1035 & 2.82, 0.60;$+$42 & 3\\
  G25.411$+$00.105 & 18 37 16.92106 & $-$06 38 30.5017 & 97.3&J1832$-$1035 & 0.35, 0.35;$+$45 & 3\\
  G26.598$-$00.024 & 18 39 55.92567 & $-$05 38 44.6424 & 24.2&J1832$-$1035 & 3.38, 0.65;$-$46 & 3\\
  G27.221$+$00.136 & 18 40 30.54608 & $-$05 01 05.3947 &118.8&J1832$-$1035 & 1.20, 0.89;$+$47 & 5\\
  G28.817$+$00.365 & 18 42 37.34797 & $-$03 29 40.9216 & 90.7&J1851$+$0035 & 4.56, 0.62;$+$41 & 4\\
  G30.318$+$00.070 & 18 46 25.02621 & $-$02 17 40.7539 & 36.1&J1851$+$0035 & 1.48, 0.71;$+$38 & 3\\
  G30.400$-$00.296 & 18 47 52.29976 & $-$02 23 16.0539 & 98.5&J1851$+$0035 & 1.36, 0.64;$+$45 & 4\\
  G31.047$+$00.356 & 18 46 43.85506 & $-$01 30 54.1551 & 80.7&J1851$+$0035 & 1.02, 0.82;$+$44 & 3\\
  G31.581$+$00.077 & 18 48 41.94108 & $-$01 10 02.5281 & 95.6&J1851$+$0035 & 0.96, 0.88;$+$50 & 2\\
  G32.992$+$00.034 & 18 51 25.58288 & $+$00 04 08.3330 & 91.8&J1851$+$0035 & 0.92, 0.82;$-$39 & 3\\
  G33.641$-$00.228 & 18 53 32.563   & $+$00 31 39.180  & 58.8&J1851$+$0035 & 1.01, 0.83;$-$57 & 5\\
  G33.980$-$00.019 & 18 53 25.01833 & $+$00 55 25.9760 & 58.9&J1851$+$0035 & 1.03, 0.80;$-$58 & 4\\
  G34.751$-$00.093 & 18 55 05.22296 & $+$01 34 36.2612 & 52.7&J1851$+$0035 & 1.02, 0.81;$-$60 & 5\\
  G35.793$-$00.175 & 18 57 16.894   & $+$02 27 57.910  & 60.7&J1851$+$0035 & 1.15, 0.80;$-$53 & 4\\
  G36.115$+$00.552 & 18 55 16.79345 & $+$03 05 05.4140 & 73.0&J1851$+$0035 & 1.46, 0.70;$-$51 & 5\\
  G36.705$+$00.096 & 18 57 59.12288 & $+$03 24 06.1124 & 53.1&J1851$+$0035 & 2.09, 0.65;$-$48 & 3\\
  G37.030$-$00.039 & 18 59 03.64233 & $+$03 37 45.0861 & 78.6&J1851$+$0035 & 2.05, 0.72;$+$42 & 6\\
  G37.598$+$00.425 & 18 58 26.79772 & $+$04 20 45.4570 & 85.8&J1851$+$0035 & 2.36, 0.64;$+$42 & 6\\
  G38.038$-$00.300 & 19 01 50.46947 & $+$04 24 18.9559 & 55.7&J1851$+$0035 & 2.74, 0.65;$-$48 & 5\\
  G38.203$-$00.067 & 19 01 18.73235 & $+$04 39 34.2938 & 79.6&J1851$+$0035 & 2.11, 0.78;$-$47 & 10\\
  G39.100$+$00.491 & 19 00 58.04036 & $+$05 42 43.9214 & 15.3&J1851$+$0035 & 2.24, 0.77;$-$46 & 11\\
  \hline           
  \multicolumn{7}{l}{$^*$ The Galactic coordinates of the brightest 6.7\,GHz methanol maser spots 
(Bartkiewicz et al. \cite{bartkiewicz09})}\\ 
\end{tabular}                                                                                
\end{table*}        

\section{Results}
The observational results are summarized in Table~\ref{table:2} and
Figure~\ref{fig1}. Table~\ref{table:2} lists the coordinates of the brightest 
water maser spot in each target, the LSR velocity (V$_{\rm p}$), and the 
intensity (S$_{\rm p}$) as well as the velocity extent of the water emission 
($\Delta$V) and the integrated flux density (S$_{\rm int}$). In most cases the
Galactic names of the water maser sources are the same as those of the
methanol masers in Bartkiewicz et al.\ (\cite{bartkiewicz09}). However, for 
five water maser sources the names are updated (marked by $^1$) as their 
positions differ by more than 3\farcs6 (0\fdg001) from the methanol maser 
positions. The two columns of Table \ref{table:2}, $\Delta_{\rm wm}$,  give 
the angular separation of two nearest spots of both
species and the corresponding difference in velocity. The last two columns,
PA$_{\rm H_2O}$ and PA$_{\rm MIR}$, list position angles of water maser
emission and MIR counterpart if it exists (Sect.~3.3). PA is defined as East
of North in the whole paper.

In Figure~\ref{fig1}, we present the spectra and angular distributions of the
water maser emission for the detected sources. The spectra were
extracted from the map data cubes using the AIPS task ISPEC and
represent the total flux density of maser emission measured in the maps. All spots
detected in each of the individual channel maps are shown. Overlaid
are the spectra and distributions of the 6.7\,GHz methanol masers as
obtained with the EVN (Bartkiewicz et al.\ \cite{bartkiewicz09}). The
parameters of all detected  water maser spots of each source are listed in
Table~\ref{table:3}.  Specifically, the position ($\Delta$RA,
$\Delta$Dec) relative to the brightest 6.7\,GHz {\it methanol} maser
spot (as listed in Table~\ref{table:1}), the LSR velocity (V$_{\rm LSR}$) 
and the intensity (S) of the maser spots are given.

Due to the relatively poor spectral resolution of 0.65\,km\,s$^{-1}$ 
of our water maser spectra we postpone an analysis of line profiles 
until follow up VLBI observations when a higher spectral resolution will be used.


\begin{figure*}
\centering
\includegraphics{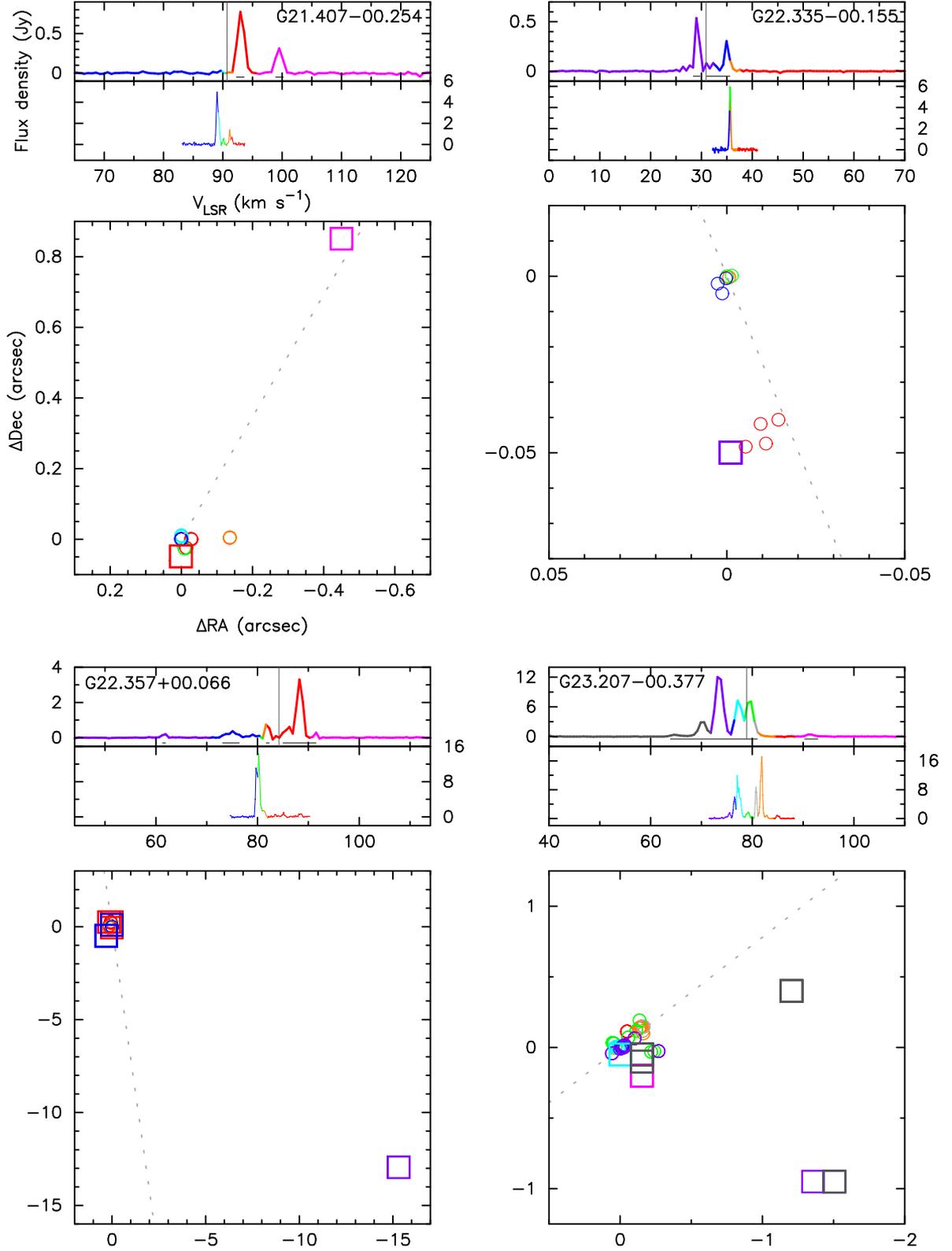}
\caption{Spectra and maps of the 22\,GHz water (VLA) and 6.7\,GHz methanol
         (EVN) maser emission. The upper and lower panels correspond to
          the water and methanol maser spectra, respectively. The thin bars under 
          the spectra show the velocity ranges of the displayed water maser spots. 
          The thin grey lines represent the systemic velocities of sources
          (Table~\ref{table:4}). Each square represents a 22\,GHz water maser spot observed in a
          single channel. 
          Note, the typical absolute positional uncertainty
          of water emission is 0\farcs15. The circles represent the 6.7\,GHz methanol
          maser spots from Bartkiewicz et al.\ (\cite{bartkiewicz09}) with
          the absolute positional accuracy of a few mas. The
          colours of squares and circles relate to the LSR velocities as
          indicated in the spectra. The coordinates are
          relative to the brightest spots of {\it methanol} emission (Table
          \ref{table:1}). Note, the source names are the Galactic 
          coordinates of the brightest spot of the {\it methanol} maser. The
          dotted lines correspond to the PA$_{\rm MIR}$ of
          4.5\,$\mu$m counterparts as listed in Table~\ref{table:2}. 
          {\it The colour version is available on-line.}}
        \addtocounter{figure}{-1}
\label{fig1}
\end{figure*}

\begin{figure*}   
\centering   
\includegraphics{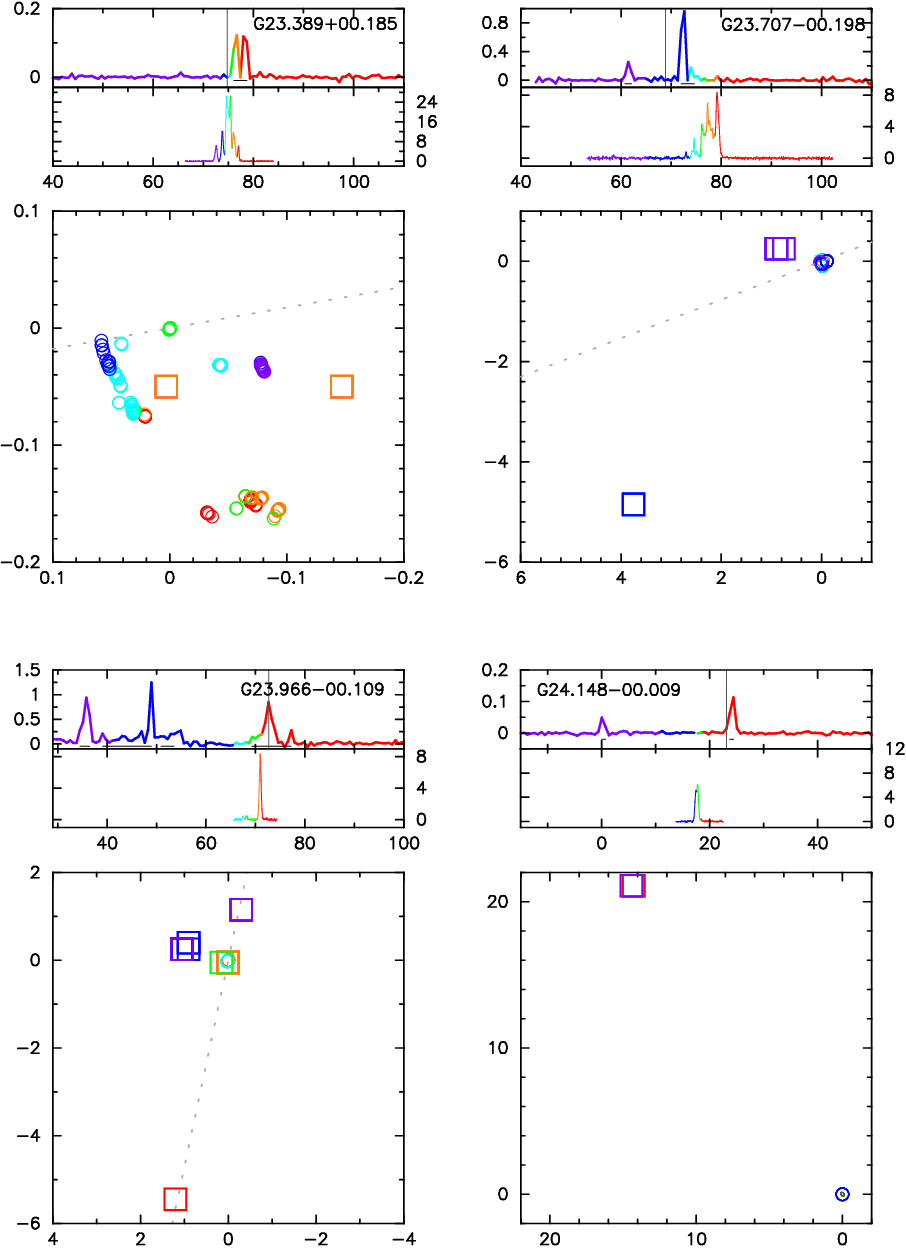}
\caption{continued. The radio continuum emission at the level of
3$\sigma_{\mathrm{rms}}$ detected toward G24.148$-$00.009 is also indicated
by a black contour (Bartkiewicz et al.~\cite{bartkiewicz09}).}
\addtocounter{figure}{-1}
\end{figure*}
\begin{figure*}   
\centering   
\includegraphics{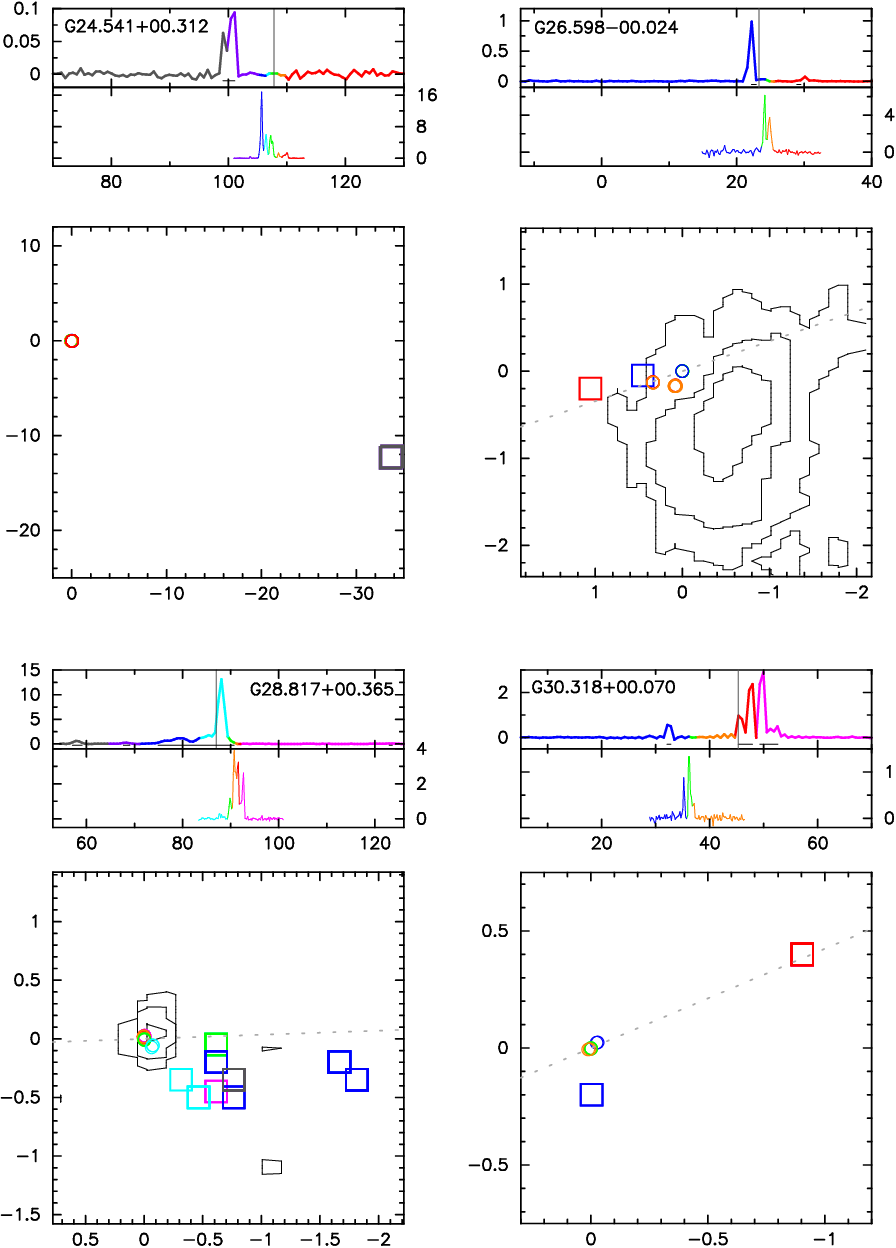}
\caption{continued. The radio continuum emission at the levels of 3, 10 and
30 $\times$ $\sigma_{\mathrm{rms}}$ detected towards
G26.598$-$00.024 and G28.817$+$00.365 are indicated by black contour lines
(Bartkiewicz et al.~\cite{bartkiewicz09}).}
\addtocounter{figure}{-1}
\end{figure*}
\begin{figure*}   
\centering   
\includegraphics{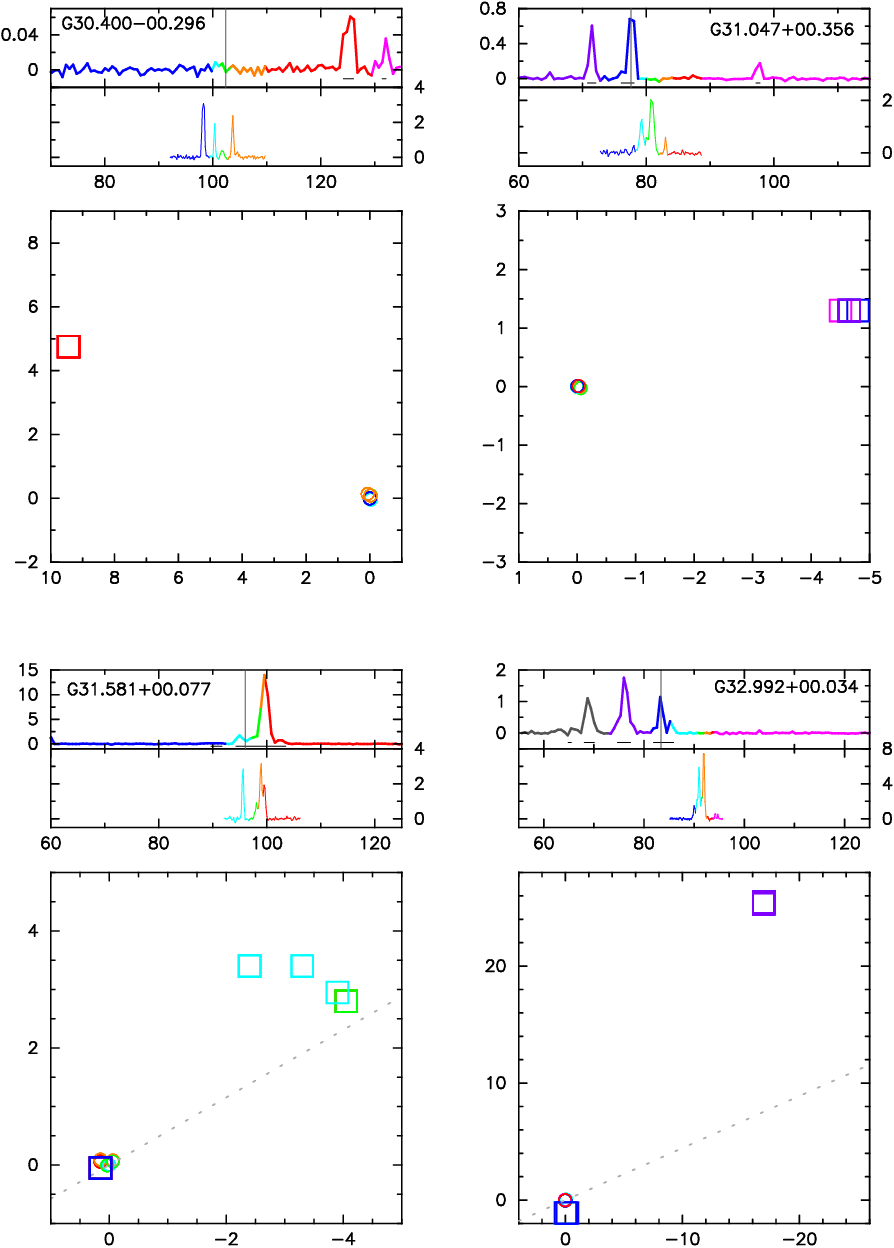}
\caption{continued.}
\addtocounter{figure}{-1}
\end{figure*}
\begin{figure*}   
\centering   
\includegraphics{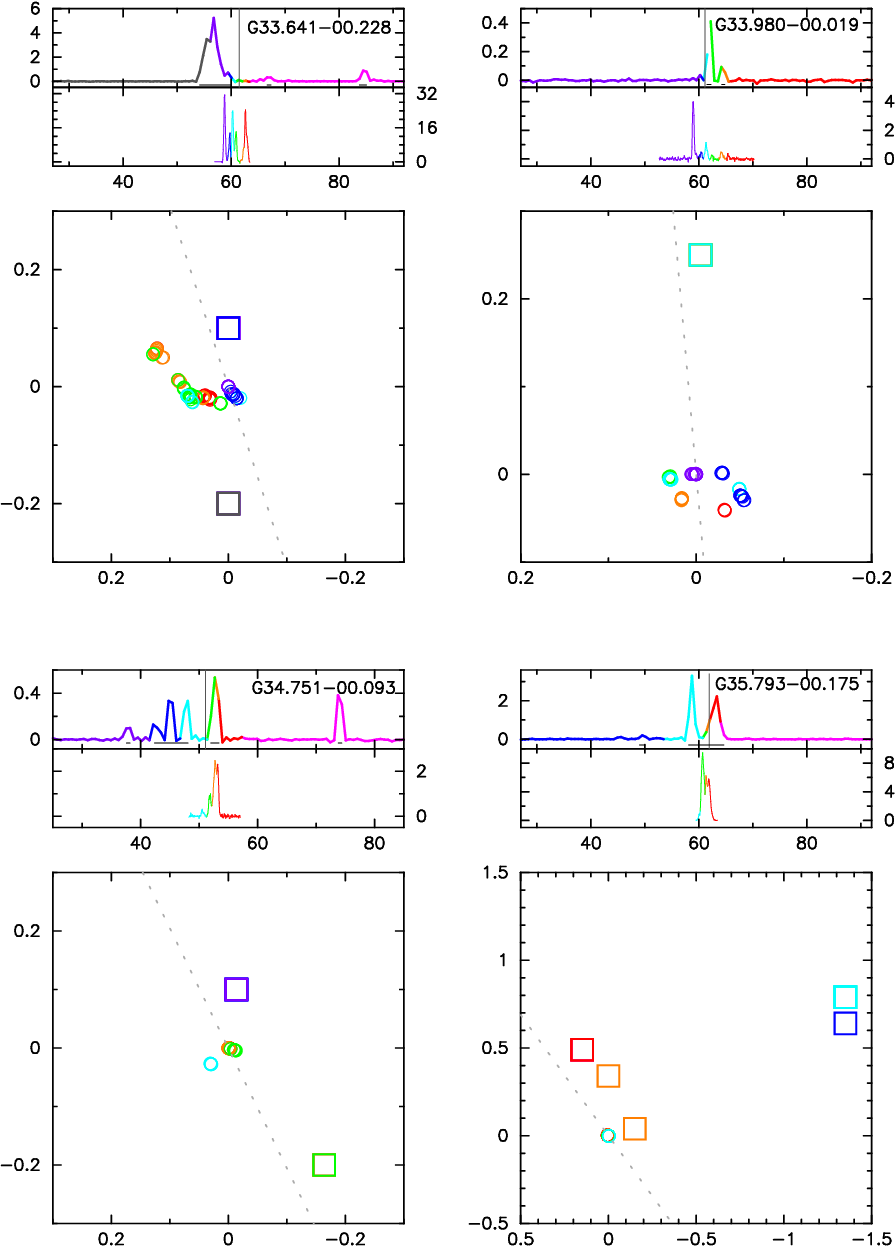}
\caption{continued.}
\addtocounter{figure}{-1}
\end{figure*}
\begin{figure*}   
\centering   
\includegraphics{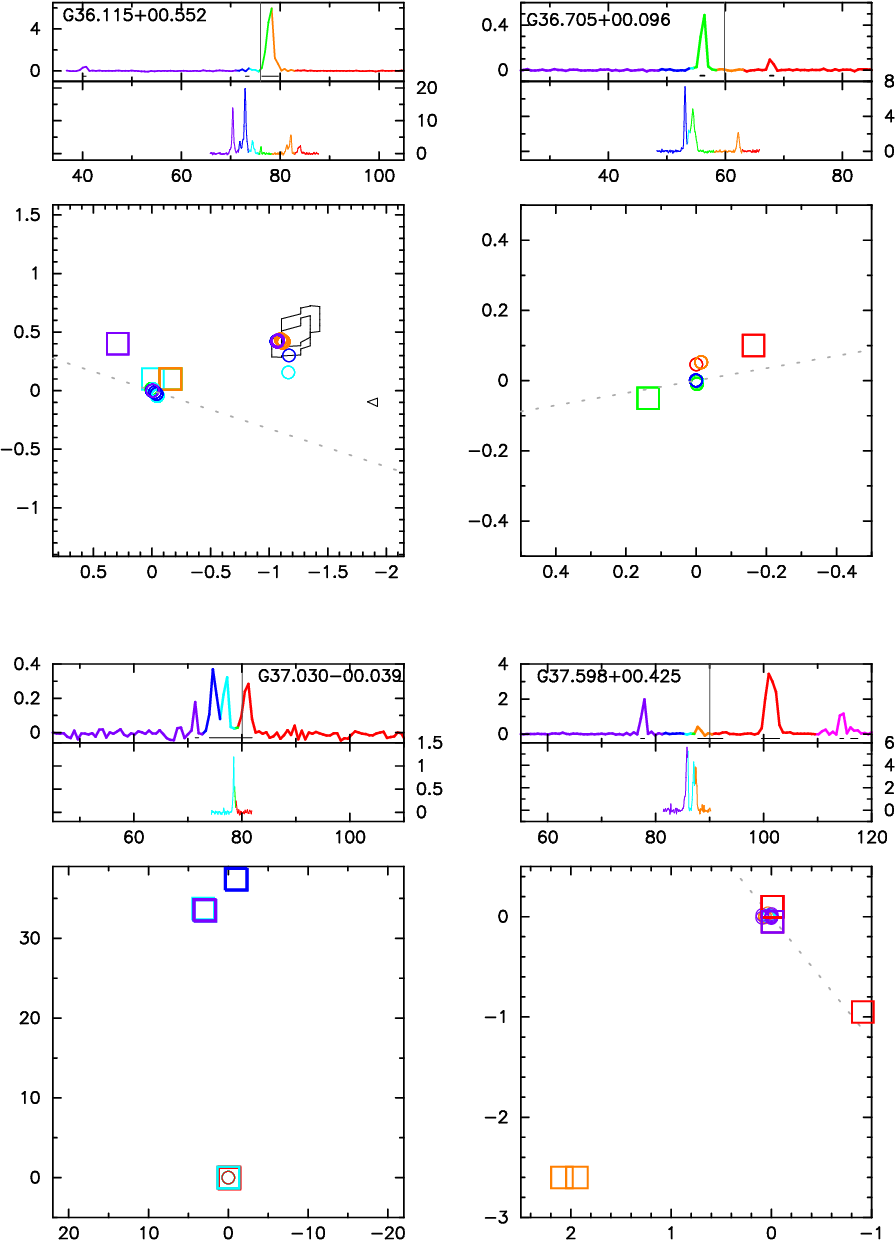}
\caption{continued. The radio continuum emission at the levels of 3 and
6 $\times$ $\sigma_{\mathrm{rms}}$ detected toward
G36.115$+$00.552 is indicated by black contour lines
(Bartkiewicz et al.~\cite{bartkiewicz09}).}
\addtocounter{figure}{-1}
\end{figure*}
\begin{figure*}   
\centering   
\includegraphics{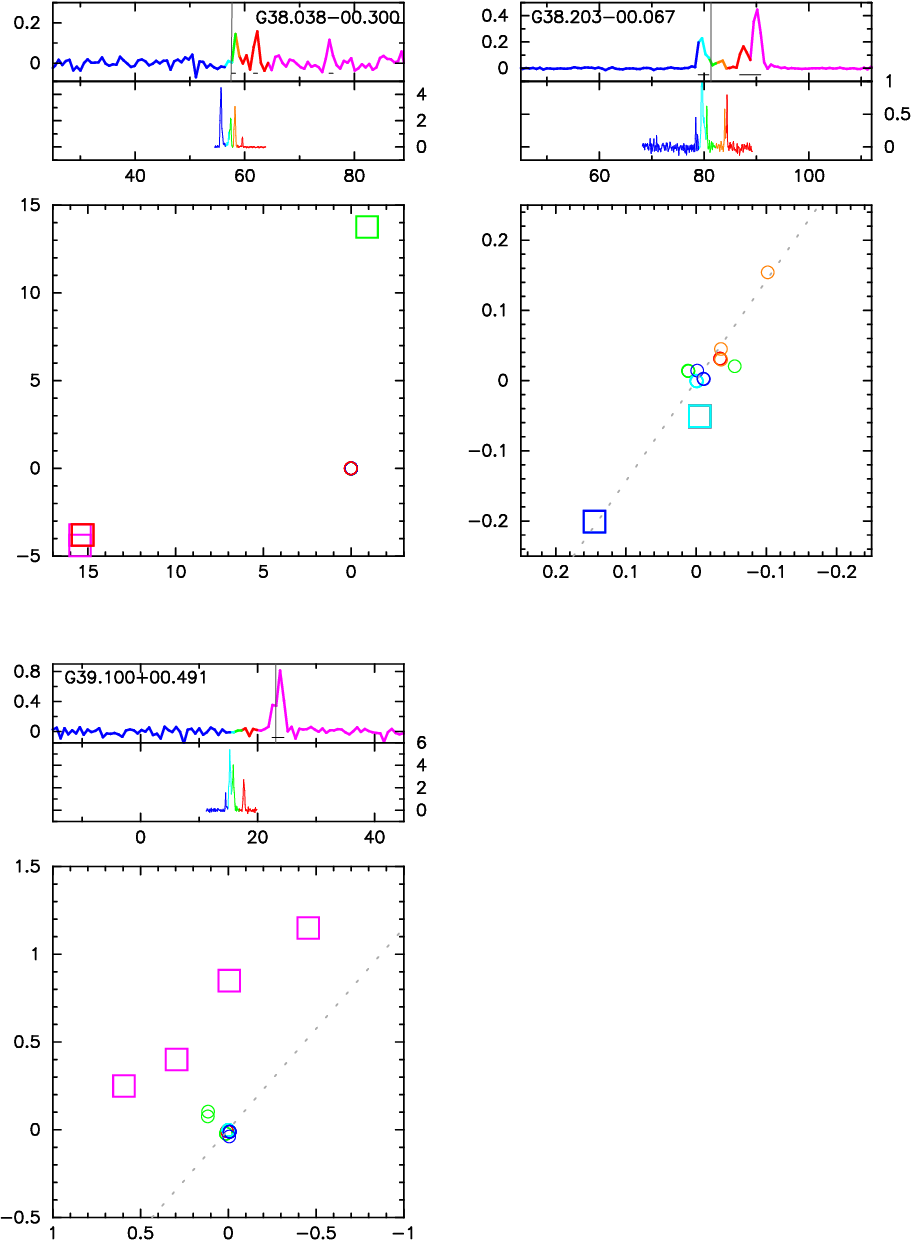}
\caption{continued.}
\end{figure*}

\begin{table*}
\caption{Results of H$_2$O observations}             
\label{table:2}      
\centering          
\begin{tabular}{l l l c c c c r r r r}     
\hline\hline
     &RA(J2000) & Dec(J2000) & V$_{\rm p}$ & $\Delta$V & S$_{\rm p}$ & S$_{\rm int}$
&\multicolumn{2}{c}{$\Delta_{\rm wm}$}& PA$_{\rm H_2O}$ & PA$_{\rm MIR}$ \\
Gll.lll$\pm$bb.bbb& (h m s) & (\degr \arcmin \arcsec)& 
(km\,s$^{-1}$) & (km\,s$^{-1}$) & (Jy\,b$^{-1}$)&
(Jy\,km\,s$^{-1}$) & (\arcsec)& (km\,s$^{-1}$) & (\degr) & (\degr) \\
\hline                    
 G21.407$-$00.254$^2$   & 18 31 06.3380 & $-$10 21 37.460 & 92.9 & 7.9 & 0.68 & 1.57 &0.03&   2.2& $-$27& {\it$-$30}\\ 
 G22.335$-$00.155$^2$   & 18 32 29.4070 & $-$09 29 29.734 & 29.0 & 7.2 & 0.76 & 1.12 &0.01&$-$2.6& 12   & 22\\
 G22.357$+$00.066       & 18 31 44.1210 & $-$09 22 12.362 & 88.3 & 30.2& 3.14 & 7.75 &0.04&   1.5& $-$29&  8\\
 G23.207$-$00.377       & 18 34 55.2019 & $-$08 49 14.943 & 73.2 & 29.0& 11.46& 55.8 &0.06&   1.2& 57   & {\it$$52}\\
 G23.389$+$00.185       & 18 33 14.3250 & $-$08 23 57.522 & 78.0 & 2.6 & 0.15 & 0.28 &0.03&$-$0.9& 90   & {\it$-$80}\\
 G23.657$-$00.127       &                &                  &      &     & $<$0.015&     &    &    &    &  \\
 G23.707$-$00.198$^2$   & 18 35 12.4165 & $-$08 17 39.108 & 72.6 & 13.8& 0.91 & 0.57 &0.77&$-$1.3& 64   & {\it$-$69}\\
 G23.966$-$00.109       & 18 35 22.2150 & $-$08 01 22.520 & 48.9 & 42.7& 1.12 & 5.98 &0.03&   1.8&$-$13 &  $-$12\\
 G24.155$-$00.010$^{1,2}$& 18 35 21.9019 & $-$07 48 34.575& 24.4& 24.4& 0.17&0.20 &25.5&  5.3 &    &    \\
 G24.534$+$00.319$^1$   & 18 34 53.4636 & $-$07 19 19.000 & 101.1& 2.0& 0.19&0.18 & 35.8&$-$2.7&   &   \\
 G24.634$-$00.324       &                &                  &      &     & $<$0.015&     &    &   &    &     \\ 
 G25.411$+$00.105       &                &                  &      &     & $<$0.015&     &    &   &    &    \\
 G26.598$-$00.024       & 18 39 55.9561 & $-$05 38 44.692 & 22.2 & 7.3 & 1.09 & 0.54 &0.14&$-$3.7& $-$76 & $-$71\\
 G27.221$+$00.136       &                 &                  &      &     & $<$0.025&     &    &  &  &    \\
 G28.817$+$00.365       & 18 42 37.2368 & $-$03 29 41.121 & 88.1 & 66.5& 13.07&34.76 &0.37&$-$2.7& $-$89 & $-$88\\
 G30.318$+$00.070       & 18 46 25.0260 & $-$02 17 40.954 & 49.9 & 20.4& 2.81 & 9.99 &0.20&$-$4.2& $-$56 & {\it $-$67}\\
 G30.403$-$00.297$^{1,2}$& 18 47 52.9295 & $-$02 23 11.303 &125.5&7.9 &0.08& 0.17   &10.5 & 19.8&  & \\
 G31.047$+$00.357$^{1,2}$& 18 46 43.5549 & $-$01 30 52.855 &77.4 & 27.0& 0.67&1.89 &4.64 & 14.8&  & \\
 G31.581$+$00.077$^2$   & 18 48 41.9510 & $-$01 10 02.578 & 99.5 & 13.8& 12.55& 2.22 &0.11 &  1.8& $-$53 &$-$60\\
 G32.992$+$00.034$^2$   & 18 51 25.5820 & $+$00 04 07.233 & 76.0 & 20.4& 1.61 & 0.54 &1.06 &$-$3.8&90 & {\it$-$66}\\
 G33.641$-$00.228       & 18 53 32.5630 & $+$00 31 39.130 & 56.8 & 30.9& 5.12 & 15.4 &0.10 &  0.7 & 0 & 18\\
 G33.980$-$00.019$^2$   & 18 53 25.0180 & $+$00 55 26.076 & 62.2 &  3.3& 0.40 & 0.49 &0.25 &  2.6 & 1 & 5 \\
 G34.751$-$00.093$^2$   & 18 55 05.2220 & $+$01 34 36.361 & 52.7 & 36.8& 0.53 & 2.51 &0.10 &$-$4.6& 27 & 26\\
 G35.793$-$00.175$^2$   & 18 57 16.8840 & $+$02 27 57.950 & 58.7 & 15.1& 2.39 & 3.53 &0.16 &  1.3 &35 & 36\\
 G36.115$+$00.552       & 18 55 16.7830 & $+$03 05 05.514 & 78.3 & 39.5& 4.91 & 10.24&0.09 &$-$2.1&57 & {\it 72}\\
 G36.705$+$00.096$^2$   & 18 57 59.1320 & $+$03 24 06.062 & 56.4 & 12.5& 0.55 & 0.71 &0.15 &  0.9& $-$64 & $-$80\\
 G37.030$-$00.039$^2$   & 18 59 03.6420 & $+$03 37 45.086 & 81.2 & 10.5& 0.27 & 0.78 &0.01 &$-$1.3 &  &   \\
 G37.598$+$00.425       & 18 58 26.7970 & $+$04 20 45.407 & 100.9& 40.2& 2.91 & 11.69&0.05 & 13.9 & 42 & {\it 39}\\
 G38.041$-$00.298$^{1,2}$& 19 01 50.4088 & $+$04 24 32.705 & 62.3 & 17.8& 0.15 & 0.28 &13.8 &  2.6 &  &\\
 G38.203$-$00.067$^2$   & 19 01 18.7320 & $+$04 39 34.243 & 90.1 & 11.9& 0.63 & 1.54 &0.05 &  0.7 &$-$45 & {\it$-$35}\\
 G39.100$+$00.491      & 19 00 58.0601 & $+$05 42 44.321 & 23.9 & 2.0 & 0.64 & 1.32 &0.35 &  8.6 & $-$48 & $-$41\\           
\hline               
\multicolumn{11}{l}{$^1$ The position of the H$_2$O maser differs by more than
3\farcs6 from that
of CH$_3$OH maser (Bartkiewicz et al.\ \cite{bartkiewicz09}) and its name is updated.} \\       
\multicolumn{11}{l}{$^2$ New detection.}\\
\end{tabular}                                                                                
\end{table*}

\subsection{Association of water and methanol masers}
In the VLA cubes of size 77$\arcsec$$\times$77$\arcsec$, water masers
were detected in 27 out of 31 cases, out of which 15 are new
detections. A total of 339 distinct maser spots were detected. To
define the detection rate of water masers actually associated with the
methanol masers, we need to determine their relative separation in physical
coordinates. The near-far distance ambiguity is not well resolved for our sources, but
it has been argued that the near kinematic distances are more likely
(Szymczak et al.\ \cite{szymczak05}). Recent measurements of
trigonometric parallaxes of several methanol sources (Reid et
al.\,\cite{reid09}; Rygl et al.\,\cite{rygl10}) strongly support this
assumption. In the following we therefore use only the near kinematic
distance estimates, and we calculated them following the prescription given by 
Reid et al.\,(\cite{reid09}). The systemic velocities, V$_{\rm sys}$, were taken either from the
observations of optically thin thermal lines (Szymczak et
al.\,\cite{szymczak07}) or from the mid-range velocity of methanol
maser features (Bartkiewicz et al.\,\cite{bartkiewicz09}). The
projected linear separation, $\Delta_{\rm wm_{dist}}$ (pc), between the nearest 
spots of the water and methanol emission were then calculated using 
the angular separation from Table~\ref{table:2}. The near 
kinematic distances for all 31 objects and the
linear separations are listed in Table~\ref{table:4}. 

\addtocounter{table}{1}
\begin{table*}
\caption{Characteristic parameters of the sources observed.}
\label{table:4}     
\centering          
\begin{tabular}{lrcrlllc}
\hline\hline
Source & V$_{\rm sys}$ & D$_{\rm near}$ &   $\Delta_{\rm
wm_{dist}}$& log($L_{\rm H_2O}$)$^{a,b}$ & log($L_{\rm H_2O}$)$^c$ & log($L_{\rm CH_3OH}$)
& Class$^1$\\
Gll.lll$\pm$bb.bbb&  (km\,s$^{-1}$)& (kpc) &  (pc)& ($L_{\sun}$)& ($L_{\sun}$) & ($L_{\sun}$) & \\
\hline                    
 G21.407$-$00.254 & 90.7 & 5.12 &  0.00074 & $-$6.02 &         & $-$5.34& C \\
 G22.335$-$00.155 & 30.9 & 2.47 &  0.00012 & $-$6.79 &         & $-$6.04& L\\
 G22.357$+$00.066 & 84.2 & 4.86 &  0.00094 & $-$5.37 & $-$7.03 & $-$5.58& C\\
 G23.207$-$00.377 & 78.9 & 4.63 &  0.00135 & $-$4.56 &         & $-$5.05& R\\
 G23.389$+$00.185 & 74.8 & 4.47 &  0.00065 & $-$6.89 &         & $-$5.14& R\\
 G23.657$-$00.127 & 82.4 & 3.19$^2$  &      & $-$8.15$\downarrow$&         & $-$5.60& R \\
 G23.707$-$00.198 & 68.9 & 4.22 &  0.01575& $-$6.63 & $-$6.34 & $-$5.13 & A\\
 G23.966$-$00.109 & 72.7 & 4.37 &  0.00063 & $-$5.57 &         & $-$5.74 & L\\
 G24.148$-$00.009 & 23.1 & 1.92 &  0.23736&$-$8.22$\downarrow$& $-$7.77 & $-$7.23 & L \\
 G24.541$+$00.312 & 107.8& 5.70 &  0.98931&$-$8.30$\downarrow$& $-$6.92 & $-$5.17 & A \\
 G24.634$-$00.324 & 42.7 & 3.00 &        &$-$8.22$\downarrow$&         & $-$6.48 & R \\
 G25.411$+$00.105 & 96.0 & 5.25 &        &$-$7.70$\downarrow$&         & $-$5.70 & R \\
 G26.598$-$00.024 & 23.3 & 1.85 &  0.00126 & $-$7.42 &         & $-$6.54 & R\\
 G27.221$+$00.136 & 112.6& 6.04 &        &$-$7.40$\downarrow$&         & $-$5.21 & C\\
 G28.817$+$00.365 & 87.0 & 4.90 &  0.00879 & $-$4.72 &         & $-$6.34 & A/R\\
 G30.318$+$00.070 & 45.3 & 2.97 &  0.00288 & $-$5.74 &         & $-$6.77 & L\\
 G30.400$-$00.296 & 102.4& 5.76 &  0.29322&$-$8.30$\downarrow$& $-$6.89 & $-$5.80 & C/R\\
 G31.047$+$00.356 & 77.6 & 4.51 &  0.10145&$-$6.05 &         & $-$6.16 &R\\
 G31.581$+$00.077 & 96.0 & 5.49 &  0.00293 & $-$4.70 &         & $-$5.82 &A/R\\
 G32.992$+$00.034 & 83.4 & 4.88 &  0.02508& $-$6.53 & $-$5.44 & $-$5.75 &C\\
 G33.641$-$00.228 & 61.5 & 3.77 &  0.00183 & $-$5.30 &         & $-$4.88 &A\\
 G33.980$-$00.019 & 61.1 & 3.75 &  0.00455 & $-$6.80 &         & $-$6.35 &R\\
 G34.751$-$00.093 & 51.1 & 3.24 &  0.00157 & $-$6.22 &         & $-$6.35 &R\\
 G35.793$-$00.175 & 61.9 & 3.83 &  0.00297 & $-$5.56 &         & $-$5.53 &L\\
 G36.115$+$00.552 & 76.0 & 4.66 &  0.00203 & $-$5.29 &         & $-$5.29 &P\\
 G36.705$+$00.096 & 59.8 & 3.75 &  0.00273 & $-$6.64 &         & $-$6.40 &C\\
 G37.030$-$00.039 & 80.1 & 5.02 &  0.00024 & $-$6.34 & $-$6.30 & $-$5.47 &S\\
 G37.598$+$00.425 & 90.0 & 6.36 &  0.00154 & $-$4.96 & $-$6.51 & $-$5.26 &C\\
 G38.038$-$00.300 & 57.5 & 3.66 &  0.24487&$-$8.00$\downarrow$&$-$7.47; $-$7.10 & $-$5.96 & C\\
 G38.203$-$00.067 & 81.3 & 5.31 &  0.00129 & $-$6.00 &         & $-$5.66 &C\\
 G39.100$+$00.491 & 23.1 & 1.70 &  0.00288 & $-$7.06 &         & $-$6.29 &C\\                    
\hline           
\multicolumn{8}{l}{$^1$ Class of morphology of {\it methanol} masers:
S -- simple, L -- linear, R -- ring, C -- complex, A -- arched,}\\
\multicolumn{8}{l}{P -pair (Bartkiewicz et al.\,\cite{bartkiewicz09}).}\\
\multicolumn{8}{l}{$^2$ Distance based on the trigonometric parallax (Bartkiewicz et al.\,\cite{bartkiewicz08}).}\\ 
\multicolumn{8}{l}{$^a$ Luminosity of water maser associated with the methanol source.
In a few cases (e.g., G22.357$+$00.066)}\\
\multicolumn{8}{l}{only some spots lie in close surrounding of methanol emission (Fig.~\ref{fig1}).}\\
\multicolumn{8}{l}{$^b$ The upper limit is marked by symbol $\downarrow$. It
means we did not register the water maser spots coinciding}\\
\multicolumn{8}{l}{ with methanol masers.}\\
\multicolumn{8}{l}{$^c$ Luminosity of water maser unassociated or likely unassociated with methanol source.} \\
\end{tabular}                                                                                
\end{table*}        

\begin{figure}
\centering
 \resizebox{\hsize}{!}{\includegraphics{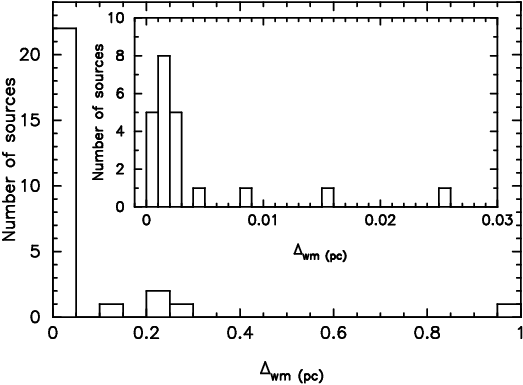}}
\caption{Histogram of linear separations between the water and methanol masers for the sample. 
   The inset is the enlargement of the histogram for the first bin.}
\label{fig2}
\end{figure}

For the majority of the detections (22 of 27), the methanol emission is
displaced by less than 0.026\,pc with a median value of 0.0017\,pc
(Table~\ref{table:4}, Fig.~\ref{fig2}). 
In these sources the velocity
difference between the nearest spots of both masers ranges from 0.7 to
13.9\,km\,s$^{-1}$, with a median value of 1.95\,km\,s$^{-1}$. The
intrinsic separation of the water and methanol spots may be slightly different  
because the position uncertainty of ~0\farcs15 results in
0.002$-$0.005\,pc displacement for our sources and there is likely an
additional spatial offset along the line of sight not accounted for
using only the projected separation. It is striking that the largest
linear separation of 0.026\,pc, for the objects considered to have
associated methanol and water masers, is consistent with the mean
separation of $\sim$0.03\,pc between the stellar objects observed in
the Orion Nebula Cluster (McCaughrean \& Stauffer\,\cite{mccaughrean94}) while the
median separation of 0.0017\,pc well agrees with mean separation of 
0.002\,pc between protostellar objects 
predicted by the merging model of massive star formation (Stahler et
al.\,\cite{stahler00}). Those above suggest an association of water and
methanol masing regions with the same protostellar object in 22 sources.  
The emissions of both maser species for the
remaining five sources shows a separation $>$0.1\,pc (see
Table~\ref{table:4}, Fig.~\ref{fig2}), suggesting the two species are
associated with separate young stellar objects within a cluster.

We conclude that at least 71\% (22/31) methanol maser sources in the
sample have associated water masers. This can be explained that  
both maser species being excited by the same underlying central object or 
 closely associated objects. This detection rate is
higher than the 52\% inferred from the 100\,m dish observation of a
much larger sample (Szymczak et al.\,\cite{szymczak05}). However, we
note that the 100\,m dish survey was about 60 times less sensitive
than the VLA observations. Considering the VLA data above a flux of 0.45\,Jy
(the mean rms noise value of observations using the Effelsberg antenna) 
we obtain a detection rate of 55\%. Our analysis
demonstrating an intrinsic association of both methanol and water masers
with the same underlying object or closely projected objects 
suggests that the two maser species
share a common stage in the early evolution of massive star. 

An inspection of the water and methanol maser spectra for the 22
objects (Fig.~\ref{fig1}), for which both types of masers are excited
by the same underlying central star, reveals that in about two-thirds of the
sources the water emission does not appear at the same velocities as
the methanol emission. In G22.335$-$0.155, G23.207$-$0.377,
G23.389$+$0.185, G31.581$+$0.077, G34.475$-$\\$-$0.093,
G38.038$-$0.300, G38.203$-$0.067 only a few features of both maser
species coincide in velocity. Furthermore, the velocity spread of the
water masers is 2$-$15 times larger than that of the methanol masers
with the exception of G23.389$+$0.185, G23.707$-$00.198,
G33.980$-$00.019, G36.705$+$0.096, G39.100$+$\\$+$00.491. That was also
found in a larger sample observed using ATCA by Breen et al.
(\cite{breen10}). This implies
that the water and methanol masers emerge from different portions of
the gas surrounding the protostar. It is consistent with theoretical
models which propose that radiative pumping of CH$_3$OH molecule
occurs at temperatures less than 150\,K and density less than
10$^8$\,cm$^{-3}$ (e.g., Cragg et al.\,\cite{cragg05} and references
therein), but the collisional pumping of H$_2$O molecules occurs in
dense ($>$10$^8$\,cm$^{-3}$) and hot (400\,K) shocked gas (Elitzur et
al.\,\cite{elitzur89}).

\subsection{Methanol sources without water emission}
Towards four sources, G23.657$-$0.127, G24.634$-$0.324,
G25.411$+$0.105, G27.221$+$0.136, no water emission was detected above a
5$\sigma$ level of 15$-$25\,mJy (Table~\ref{table:2}). Three of them
(G23.657$-$0.127, G24.634$-$0.324 and G25.411$+$0.105) show a ring-like
structure of the 6.7\,GHz methanol maser emission (Bartkiewicz et
al.\,\cite{bartkiewicz09}). Such morphologies have been found recently 
in at least nine out 31 sources (Bartkiewicz et
al.\,\cite{bartkiewicz09}) and was the reason for these follow-up
observations. In addition there are five methanol
sources (G24.148$-$00.009, G24.541$+$00.312, G30.400$-$00.296,
G31.047$+$00.356 and G38.038$-$00.300) where the water masers seem to be 
unassociated since the linear distance between both masers is above 0.1\,pc 
(Table~\ref{table:4}). Their methanol masers have linear, arched, complex/ring,
ring and complex structures, respectively (Bartkiewicz et 
al.\,\cite{bartkiewicz09}). For the clarity we list the morphological
classification of all {\it methanol} masers in the last column in 
Table~\ref{table:4}. It is interesting
that towards G38.038$-$0.300 two distinct water masers were detected, 
but both were offset about 15", corresponding to 0.2\,pc separation for the
near kinematic distance. 

We note that 
emission from the 22\,GHz water transition often exhibits significant
temporal variability on time scales of a few months (Brand et
al.\,\cite{brand03}). Therefore we may expect that a number of
non-detections in our sample can be different at other epoch. 
However, our non-detections were also not detected by the single
dish study (Szymczak et al.\,\cite{szymczak05}), when observed with a
sensitivity of $\sim$1.5\,Jy.  Thus, the water emission in these
sources may be relatively weak if present at all.

\subsection{Maser luminosity}

\begin{figure}
\centering
 \resizebox{\hsize}{!}{\includegraphics{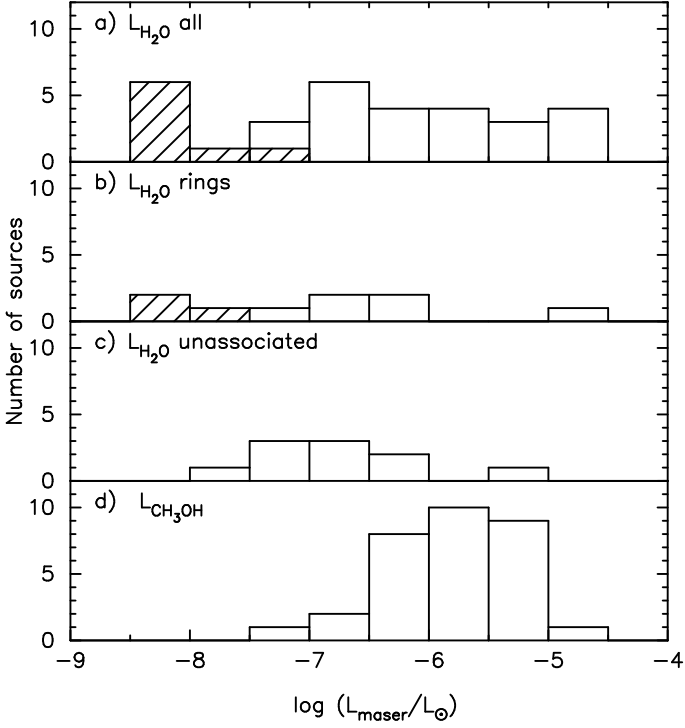}}
 \caption{{\bf a)} Histogram of the water maser luminosity in the
   6.7\,GHz methanol maser sample (Table~\ref{table:4}. 
   The dashed bars mark sources with the upper luminosity limits (non
   detections). {\bf b)} Same as in {\bf a)} but for
   the sources with ring-like distribution of methanol emission.
   {\bf c)} Same as in {\bf a)} but for the water maser sources not
   associated with the methanol masers.  {\bf d)} Histograms of
   methanol maser luminosity in the sample.}
\label{fig5}
\end{figure}
We have calculated the maser luminosities using the VLA data for the water 
maser emission and the 32\,m dish observations for the methanol maser emission
(Table~\ref{table:4}). They were observed almost simultaneously (Sect.~2.2). 
In our sample, the {\it isotropic} water maser luminosity ranges from 10$^{-7.4}$ to
10$^{-4.6}L_{\sun}$. The median luminosity for the whole sample is
10$^{-6}L_{\sun}$. We note, that all but one, G23.207$-$00.377,  
sources with the ring morphology of methanol emission have water maser luminosities 
{\it lower} than 10$^{-6}L_{\sun}$ (Fig.~\ref{fig5}).  
This suggests that these sources are associated with young massive
stellar objects, in which water masers are less luminous than in the
methanol sources of other morphology. 

The water masers detected in our survey that are not associated with
methanol masers have a median H$_2$O maser luminosity of
10$^{-6.9}L_{\sun}$ (Fig.~\ref{fig5}). Thus they do not significantly
differ from the luminosity of water masers that are associated with
methanol sources. 
The median luminosity of methanol maser emission, estimated from
single dish data, is 10$^{-5.7}L_{\sun}$, somewhat higher than
that of the water maser emission (Fig.~\ref{fig5}). We relate this difference to the
extremely high sensitivity of the VLA observations, about 60 times
better than that reported in Szymczak et al. (\cite{szymczak05}). We do
not notice any correlation between luminosities of methanol and water
masers. Xu et al. (\cite{xu08}) found a correlation between these both
values, however they claimed that since there were no physical
connections between both lines, it might be a distance squared effect, 
as suggested by Palla et al.~(\cite{palla91}).  
The two maser species require different excitation conditions
and, even if they are related to the same YSO (Sect.~3.1), may arise in different
subregions such as discs and outflows. Further, Xu et al.~(\cite{xu08})
found water maser luminosities were higher than the methanol maser
luminosities, a finding that is not supported by our data.

\section{Discussion}
\subsection{Morphologies of masers and MIR counterparts}
We are interested in studying the
morphology of the masers in relation to that of the dust. 
So, we searched for mid-infrared (MIR) emission toward the detected  water
masers using the {\it Spitzer} IRAC 
maps\footnote{http://irsa.ipac.caltech.edu/}. In total we
found that 21 out 27 sources of water maser emission have MIR counterparts
within 1\farcs2 arcseconds on the sky
(Table~\ref{table:2}). To clarify the nature of the studied sources we compare the
morphology of the water maser and MIR emission from maps at 4.5\,$\mu$m of pixel size
of 0\farcs6 (Fazio et al.\,\cite{fazio04}). 
The position angle of water maser clusters associated with the 
methanol source, PA$_{\rm H_2O}$, was determined using a least square 
fit to the maser spot
distribution. For sources with a single water maser cluster the PA$_{\rm
  H_2O}$ was assumed to be a position angle of the direction between
the water maser cluster and the flux-weighted centre of the 6.7\,GHz
methanol maser distribution observed with the EVN (Bartkiewicz et
al.\,\cite{bartkiewicz09}). The PA$_{\rm MIR}$ was estimated using 
4.5\,$\mu$m emission maps and by fitting
two-dimensional Gaussian components. For several objects these
estimates are very uncertain, and instead we used the maps of the
4.5\,$\mu$m$-$3.6\,$\mu$m excess. The values of PA$_{\rm H_2O}$ and
PA$_{\rm MIR}$ are listed in the two last columns of
Table~\ref{table:2}. The entries with an error of about 10$-$15\degr\,
are given in italic. For the remaining sources errors in PA$_{\rm
  MIR}$ and PA$_{\rm H_2O}$ are smaller than 6\degr\, and 3\degr,
respectively.
  
\begin{figure}
\centering
 \resizebox{\hsize}{!}{\includegraphics{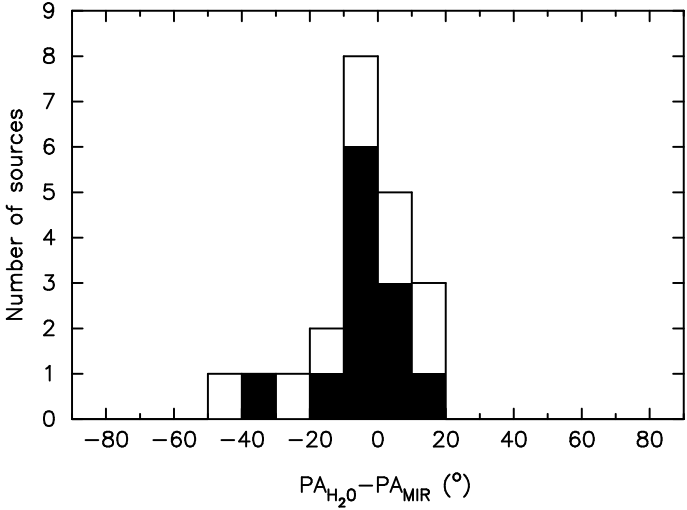}}
 \caption{Histogram of the differences in the position angles of major
   axes of water maser distribution and 4.5\,$\mu$m emission excess
   for the studied objects. The sources with small errors of PA$_{\rm
     MIR}$ are marked solid.}
\label{fig3}
\end{figure}

Fig.~\ref{fig3} shows the distribution of the position angle
differences. In 18 out of 21 sources the water maser structure is
aligned within less than 20\degr\, with the extended emission at
4.5\,$\mu$m. For the remaining sources the position angle differences
are less than 47\degr. As the 4.5\,$\mu$m emission is interpreted as
tracer of shocked H$_2$ in the outflow (Smith et al.\,\cite{smith06};
Davis et al.\,\cite{davis07}; Cyganowski et al.\,\cite{cyganowski08,
  cyganowski09}, and references therein) our finding of tight
alignment of the spatial extent of these two tracers strongly suggests
that the H$_2$O masers originate in outflows. It is fully consistent
with the theoretical model that H$_2$O masers are excited due to
collisional pumping with H$_2$ molecules in shocks associated with
outflows (Elitzur et al.\,\cite{elitzur89}).

Comparison of the H$_2$O maser morphology taken with the VLA-CnB with
that of 6.7\,GHz CH$_3$OH maser seen with the EVN (Bartkiewicz et
al.\,\cite{bartkiewicz09}) is difficult because of larger positional
uncertainty in the VLA 22 GHz data, and because a significant fraction
of the 6.7\,GHz flux may be missed in the milliarcsecond (mas)
resolution observations. The case of G39.100+0.491 is instructive in
this context; this relatively nearby source (distance 1.7\,kpc)
observed at 6.7\,GHz with the 5$\times$15\,mas$^2$ EVN beam appeared
as an irregular cluster of size of 0\farcs18$\times$0\farcs04
(Bartkiewicz et al.\,\cite{bartkiewicz09}). However, using the
2\farcs4$\times$1\farcs3 VLA beam, a much richer structure of methanol maser emission 
was revealed: two bright clusters separated by $\sim$0\farcs7 at PA of
$-$50\degr\, and diffuse emission between them (Cyganowski et
al.\,\cite{cyganowski09}). The 6.7\,GHz maser emission is clearly
extended along the same position angle as the 4.5\,$\mu$m emission in
the {\it Spitzer} maps, as well as the H$_2$O maser emission observed
here.

However, for sources with a ring-like distribution of
6.7\,GHz maser emission seen at mas resolution, the comparison with the
structure of water emission obtained with arcsecond resolution is still useful. 
In four out of five rings, where a methanol-water maser association exists, 
the major axis of the methanol ring is crudely orthogonal to
the main axis of the water maser structure. 
G23.207$-$0.377 is likely the best example. The part of methanol masers form a ring-like
structure with the PA of the major axis of $-$60${\degr}$, and with a
velocity range of 13.20\,km\,s$^{-1}$, the rest likely trace an outflow 
(Bartkiewicz et al.\,\cite{bartkiewicz09}). The water maser emission is distributed over a
region of 1\farcs5$\times$1\farcs5 and with the wider velocity range of
29\,km\,s$^{-1}$. The linear size of water maser is 0.04\,pc, while
the methanol ring has diameter of only 0.006\,pc. As the methanol
emission in ring-like structure likely traces a disc or torus around a
massive stellar object (Bartkiewicz et al.\,\cite{bartkiewicz09}) the
water maser emission along the normal to the disc implies the
outflow. This is supported by MIR counterpart which shows PA of
52${\degr}$ (Table~\ref{table:2}). The other three sources,
G23.389$+$0.185, G33.980$-$0.019 and G34.751$-$0.093 are possible
cases with a  similar scenario.


We conclude that the water masers in our sample generally coincide with the MIR
counterpart sources. The majority of the sources show a water maser
structure aligned with the extension direction of the 4.5\,$\mu$m
emission, while in particular for some of the objects, where the
methanol masers are tracing circumstellar discs/tori, the water masers
appear in the orthogonal direction. In general this is consistent with
previous observations showing that MIR emission is indeed associated
with water masers, and their relative distributions indicate that
water masers originate in outflows (e.g., De Buizer et
al.\,\cite{debuizer05}).

Assuming that the methanol emission arises close to the protostar, while H$_2$O
masers trace outflows further from the central object, the size
of the outflows can be estimated in the case of 22 sources where a
methanol--water maser association exists. Our data imply size
scales for the outflows from 0.0006 to 0.13\,pc with a median and mean
values of 0.01 and 0.022\,pc, respectively. We note, that three sources: 
G22.335$-$0.155, G30.400$-$0.296 and G33.980$-$00.019 are unresolved
with a 1\farcs4$\times$0\farcs8 beam. Their velocity extents are 
from 3.3 to 7.9\,km\,s$^{-1}$ (Table~\ref{table:3}). That also may 
suggest the presence of outflows along the line of sight. However, we need
to verify that hypothesis with VLBI observations.

\subsection{Signposts of multiple active centres}
Towards 12 of the methanol targets, water maser emission was
detected in distinct clusters separated from methanol maser spots by $>$0.1\,pc. In
two objects, G37.030$-$00.039 and G38.038$-$00.300, even two water clusters
significantly displaced were seen. Therefore, in total 14 water maser
clusters were found in our sample lying significantly further from methanol 
maser spots (Table~\ref{table:5}). Such characteristic was also noticed in 
studies by Breen et al. (\cite{breen10}) using ATCA.

For these objects we also inspected {\it Spitzer}
IRAC maps to search for MIR counterparts of the water maser clusters. 
The results are summarized in Table~\ref{table:5}. We found that 10 of the 
14 clusters have MIR counterparts which, within measurement
uncertainty, do not coincide with the methanol maser clusters and their MIR
counterparts that were reported in Bartkiewicz et al.
(\cite{bartkiewicz09}). Therefore, we conclude
that in these 10 cases both masers, methanol and water, are not associated with 
the same MIR object. It is possible that 
 both masers are associated with the same molecular cloud where star 
formation takes place, but do not trace the environment of the same protostar. 
They likely trace different protostars as most of
them have MIR properties typical of embedded young massive objects
and show a 4.5\,$\mu$m$-$3.6\,$\mu$m excess that is an indicator of
shocked material in outflows (e.g., Cyganowski et al.\,\cite{cyganowski08}). 
The remaining four water maser clusters lying 0.1--0.3\,pc from the methanol
maser spots (G23.966$-$00.109, G31.581$+$00.077, G37.598$+$00.425
and tentatively G31.047$+$00.356) (Table~\ref{table:5}) are likely associated with the 
same MIR objects as the methanol emission. 

\begin{table*}
\caption{List of methanol maser sources with the water maser emission of linear offset greater than 0.1\,pc.
  The two first columns list the galactic coordinates of methanol and water masers, respectively. 
  $\Delta_{\rm wm_{dist}}$ is the angular (Col.~3) and linear (Col.~4) separation of H$_2$O maser emission from 
  the methanol source. $\Delta$V$_{\rm min}$ is the minimum and $\Delta$V$_{\rm max}$ is the maximum differences 
  between the LSR velocity of water maser spot from the analysed group and the systemic velocity
  (Table~\ref{table:4}). The name of MIR source nearby to the H$_2$O 
  maser emission and angular separation between them, $\Delta$(MIR$-$H$_2$O), are given.}
\label{table:5}     
\centering          
\begin{tabular}{l l l l l l  l l}
\hline\hline
CH$_3$OH source & H$_2$O emission & \multicolumn{2}{c}{$\Delta_{\rm
wm_{dist}}$} &$\Delta$V$_{\rm min}$ & 
$\Delta$V$_{\rm max}$ & MIR source & $\Delta$(MIR$-$H$_2$O) \\
Gll.lll$\pm$bb.bbb&Gll.lll$\pm$bb.bbb &(arcsec)&(pc)&(km s$^{-1}$)& (km s$^{-1}$)& & (arcsec) \\
\hline                    
 G22.357$+$00.066 & G22.351$+$0.068 & 20.05 & 0.47    &22.3    & 22.9  &  G022.3506+00.0678 & 3.2$^2$  \\
 G23.707$-$00.198 & G23.706$-$0.200 &  6.13 & 0.13    &3.1 &  5.0  &  G023.7057-00.1999 & 1.3$^2$  \\
 G23.966$-$00.109 & G23.965$-$0.110 & 5.58 & 0.12    &4.5$^*$    & &  G023.9649-00.1104 & 0.6$^2$  \\
 G24.148$-$00.009 & G24.155$-$0.010 & 25.48 & 0.24    &0.6 & 23.1  & G024.1550-00.0119 & 7.3$^4$  \\
 G24.541$+$00.312 & G24.534$+$0.319 & 35.90 & 0.99    &6.7 &  8.7  & G024.5351+00.3190 & 5.1$^3$ \\
 G30.400$-$00.296 & G30.403$-$0.297 & 10.57 & 0.29    &21.8& 29.7  & G030.4010-00.2960 & 7.9$^2$  \\
 G31.047$+$00.356 & G31.047$+$0.357 & 6.91 & 0.11    &0.2 & 20.2  &  G031.0467+00.3574 & $^5$  \\
 G31.581$+$00.077 & G31.581$+$0.078 & 4.92 & 0.13    &1.7 & 4.9  &  G031.5813+00.0788 &  2.3$^1$  \\
 G32.992$+$00.034 & G32.996$+$0.041 & 30.58 & 0.72    &0.2 & 17.9  & G032.9962+00.0414 & 0.3$^2$  \\
 G37.030$-$00.039 & G37.039$-$0.035 & 33.73 & 0.82    &2.8 &  8.7  & G037.0385-00.0350 & 1.6$^2$  \\
                  & G37.039$-$0.034 & 37.36 & 0.91    &4.8 &  6.1  &G037.0385-00.0350& 6.1$^2$  \\
 G37.598$+$00.425 & G37.597$+$0.424 &  3.34 & 0.10    &1.6 & 2.2  & G037.5978+00.4253 & 4.1$^1$  \\
 G38.038$-$00.300 & G38.038$-$0.305 & 16.06 & 0.28    &4.1 & 18.6  &  G038.0384-00.3042& 1.6$^2$  \\
                  & G38.041$-$0.298 & 13.78 & 0.24    &0.8$^*$ &       & G038.0409-00.2968 & 5.2$^2$  \\
\hline           
\multicolumn{8}{l}{$^1$ H$_2$O maser emission is {\bf associated} with the same MIR source as CH$_3$OH maser
emission.}\\
\multicolumn{8}{l}{$^2$ H$_2$O maser emission is {\bf not associated} with the same MIR source as CH$_3$OH maser
emission.}\\ 
\multicolumn{8}{l}{$^3$ H$_2$O maser emission is {\bf not associated} with the same MIR source as CH$_3$OH maser emission, 
            instead it lies in a cluster of three MIR sources.}\\
\multicolumn{8}{l}{$^4$ H$_2$O maser emission is {\bf likely not associated} with the same MIR source as 
            CH$_3$OH maser emission but with MIR object in the region}\\
\multicolumn{8}{l}{of diffuse ($\sim$10\arcsec) excess of 4.5\,$\mu$m emission seen in the IRAC Spitzer maps.} \\
\multicolumn{8}{l}{The name of the strongest MIR counterpart is given.} \\
\multicolumn{8}{l}{$^5$ H$_2$O maser emission is tentatively associated with the same MIR source as CH$_3$OH maser 
            emission, laying at edge of diffuse excess of }\\
\multicolumn{8}{l}{4.5\,$\mu$m emission from possible cluster of MIR sources.}\\
\multicolumn{8}{l}{$^*$ When single maser spot was observed only $\Delta$V$_{\rm min}$ is
given.}\\
\end{tabular}                                                                                
\end{table*}        

\subsection{Association and non-association with H\,{\small II} regions}
In our previous VLA project we searched for the 8.4\,GHz continuum emission 
toward the presented methanol masers with the sensitivity level of 0.15\,mJy\,beam$^{-1}$ 
(Bartkiewicz et al.\,\cite{bartkiewicz09}). In total, we detected eight
sources toward the sample. Only in four cases the continuum emission was 
associated with the methanol sources. Concerning the project presented in this paper we note that 
in three of these four objects the methanol structures are also associated with the distribution
of water masers (Fig.~\ref{fig1}). Therefore, both masers are likely physically 
connected with radio continuum emission. One example is G28.817$+$0.365 
where methanol spots are distributed along a PA 
of $+$45${\degr}$. They are projected on the central part
of a H\,{\small II} region. Water 
masers are distributed along a PA of $-$89${\degr}$ and are spread in the SW 
direction from the radio continuum source. Such a 
distribution of masers may indicate an outflow scenario for both masers. But
proper motion studies are needed to verify that scenario as for example 
presented for 12\,GHz methanol masers 
in Moscadelli et al. (\cite{moscadelli02}). We note that the MIR
counterpart is aligned with PA of $-$88${\degr}$ (Table~\ref{table:2}) and
supports the outflow hypothesis. \\
Two other sources with detected H\,{\small II} regions, G26.598$-$00.024 and 
G36.115$+$00.552, show similar characteristic. Here, the overall distributions of 
methanol and water masers are crudely aligned in
the same direction as the MIR counterparts (Table~\ref{table:2}). 
In these cases masers are displaced from the centres of the radio continuum 
sources by 0.007 and 0.02\,pc, respectively. We suggest that these three 
objects support the hypothesis that we have HMSFRs at
different evolutionary stage in our sample. Beuther \& Shepherd
(\cite{beuther05}) presented a scenario for the evolution of massive outflows.
When a B star forms via accretion through a disk and the H\,{\small II} region 
is not yet formed, the disk-outflow interaction produces a collimated outflow. 
With time a hyper-compact H\,{\small II} region forms and the wind from the 
massive young star produces a less collimated outflow. The disk begins to be 
destroyed and
an outflow with a small degree of collimation dominates the system. In 
sources with methanol maser rings, we either did not detected water maser
emission or relatively weak one. In sources with H\,{\small II} counterparts
the methanol maser morphology is less regular (although G26.598$-$00.024 was
classified as ring, we note it consists only three cluster of masers), 
and water-methanol maser spot distributions can also be 
consistent with an outflow scenario (Fig.~\ref{fig1}). 

In eight out nine objects, where no water was found (Sect.~3.2), no
continuum emission at 8.4\,GHz above $\sim$0.15\,mJy\,beam$^{-1}$ was
detected. The one exception is G24.148$-$00.009 which has weak and compact emission 
(S$_{\rm p}$=1.05\,mJy\,beam$^{-1}$; S$_{\rm int}$=1\,mJy; 
Bartkiewicz et al.\,\cite{bartkiewicz09}). We also note, that towards 
that source we did not detect any water maser within 0.24\,pc. 
That is opposite to the result obtained by Beuther et al.\
(\cite{beuther02}), where methanol masers are associated with cm
emission only if there are nearby water masers. However, as noted by
these authors, in the archetypical star forming region W3(OH) a
situation similar to that of G24.148$-$00.009 exists, where the
methanol masers are associated with an ultracompact H\,{\small II}
region and the water masers are offset significantly and associated 
with a different young star (Menten \cite{menten96}).

The absence of continuum emission and water masers may lend
support to the hypothesis that 6.7\,GHz methanol masers trace an earlier
evolutionary phase than water masers where no outflows have started yet (e.g., 
Ellingsen et al.\,\cite{ellingsen07}). For the ($\approx$50\% of)
methanol masers not associated with H\,{\small II} regions 
 (and water emission) that show ring-like structures, we suggest that they   
 trace circumstellar discs/tori, probably at an early stage of evolution. 
A less regular methanol morphology would appear in later stages and
could be related with outflows traced by water masers. Comparative
studies of water and methanol masers in the giant molecular cloud
G333.6$-$0.2 have suggested a similar conclusion that 6.7\,GHz
methanol masers trace an earlier evolutionary phase of high-mass star
formation than do luminous water masers (Breen et al.\,\cite{breen07};
Ellingsen et al.\,\cite{ellingsen07}). It is interesting that in 
four of five 
ring-like methanol sources, the associated water masers are weaker than
the methanol masers. This supports the above mentioned scenario about an early
evolutionary stage for these sources, as they may be
examples  where the outflows have just begun. Only one ring-like source
(G23.207$-$0.377) does not follow this trend, since its
methanol emission is much weaker than the water emission. However, we note 
that in the distribution of methanol maser spots in this source one can see components 
that may belong to the outflow part traced also by water emission
(Fig.~\ref{fig1}). Therefore, that object is likely more evolved. 
We present this source in more details in Sect.~3.4. 

The aforementioned interpretation of evolutionary stage relies basically on the lack of
detectable continuum at centimeter wavelengths. This may be misleading
since only the most compact continuum emission was mapped. We note, that 
in a case of G22.357$+$0.066 observations with the VLA at 
8.4\,GHz with a synthesized beam of 2\farcs3 and a sensitivity of
0.3\,mJy\,beam$^{-1}$, revealed a complex and extended source
with a peak flux of 1.02\,mJy\,beam$^{-1}$ which is 3$\sigma$ detection (van der Walt et
al.\,\cite{walt03}). However, no emission was found at the same frequency
with a beam of 0\farcs35$\times$0\farcs25 and a sensitivity of
0.15\,mJy\,beam$^{-1}$ (Bartkiewicz et al.\,\cite{bartkiewicz09}).
The methanol emission is offset by $\sim$15\arcsec\, from the radio
continuum peak, corresponding to a projected linear distance of
0.35\,pc. We therefore suggest that further sensitive searches for
millimeter and centimeter continuum counterparts of ring-like methanol
sources will be important to understand their nature.


\section{Conclusions}
High sensitivity VLA observations of the 22\,GHz water maser line
towards 31 methanol maser objects have yielded 27 detections, out of
which 15 were detected for the first time. Most (71\%) of the methanol
sources have water masers  with a projected separation of less than 0.026\,pc.
They are either excited by the same underlying central object or come from
different, but closely projected YSOs. 
We identified MIR counterparts of 21 water masers from {\it Spitzer} IRAC
maps. The water maser structures are well aligned with the extended emission at
4.5\,$\mu$m for a large fraction (18/21) of the studied objects. This
confirms that the water masers originate in outflows. 

A distinct group of sources with ring-like methanol maser distribution, likely tracing
circumstellar disc/torus around high-mass young stellar objects, show
either no associated water masers at all (4 out 9), or a water maser
distribution which is orthogonal to the major axis of the methanol
ring (4 out 9). Moreover, the majority of this group of objects (8 out 9) 
does not show detectable continuum emission at 8.4\,GHz and may 
represent an early phase of evolution.  One methanol ring,
G26.598$-$00.024, lies at the edge of a H\,{\small II} region and is aligned
with associated water masers. Both masers water and methanol masers likely form in outflows. 

We suggest that massive star forming regions that contain methanol masers with
ring-like morphologies are at the earliest evolutionary states when the young 
star is still forming, possibly via accretion through a disk. When winds begin to
dominate, the regular ring-like structure is destroyed and methanol and water 
masers appear to be associated with the outflow. 
Further deep observations of these sources in the radio continuum as well as 
in the infrared range are required to explain their nature.

\begin{acknowledgements}
A.B.\ and M.S.\ acknowledge support by the Polish Ministry of Science and 
Higher Education through grant N N203 386937. A.B.\ acknowledges support by
the Nicolaus Copernicus University grant 364-A (2009).\\
The Very Large Array (VLA) of the National Radio Astronomy Observatory is a    
facility of the National Science Foundation operated under cooperative
agreement by Associated Universities, Inc. This research has made use
of the NASA/IPAC Infrared Science Archive, which is operated by the   
Jet Propulsion Laboratory, California Institute of Technology, under  
contract with the National Aeronautics and Space Administration.

\end{acknowledgements}

\clearpage
\onecolumn
\setcounter{table}{2}
\begin{table}
\caption{Water maser spots detected towards the sample of 31 methanol maser sources. The
absolute coordinates of the (0,0) point for each object are listed in Table~\ref{table:1}.}
\label{table:3}      
\centering          
\begin{tabular}{rccr}     
\hline\hline       
\\
$\Delta$RA & $\Delta$Dec & V$_{\rm LSR}$ & S \\
(\arcsec)     &  (\arcsec)      & (km\,s$^{-1}$)& (mJy\,beam$^{-1}$)\\
\hline
\multicolumn{4}{l}{\bf G21.407$-$00.254}\\
       -0.450&        0.851&             100.2&     151.470\\
       -0.450&        0.851&              99.5&     290.950\\
       -0.450&        0.851&              98.9&     170.910\\
        0.001&       -0.049&              93.6&     457.630\\
        0.001&       -0.049&              92.9&     681.440\\
        0.001&       -0.049&              92.3&     319.830\\
\multicolumn{4}{l}{\bf G22.335$-$00.155}\\
       -0.001&       -0.050&              35.6&     169.200\\
       -0.001&       -0.050&              34.9&     425.210\\
       -0.001&       -0.050&              32.3&     112.560\\
       -0.001&       -0.050&              31.0&     112.750\\
       -0.001&       -0.050&              29.7&     402.000\\
       -0.001&       -0.050&              29.0&     764.310\\
       -0.001&       -0.050&              28.4&      55.741\\
\multicolumn{4}{l}{\bf G22.357$+$00.066}\\                                                       
        0.007&        0.101&              91.5&     268.630\\
        0.007&        0.101&              89.6&      97.639\\
        0.157&        0.251&              88.9&    1935.500\\
        0.007&        0.251&              88.3&    3136.500\\
        0.007&        0.101&              87.6&    1570.100\\
        0.157&        0.251&              86.9&     248.040\\
        0.007&        0.101&              86.3&     578.120\\
        0.007&        0.101&              85.6&     364.780\\
        0.007&        0.101&              85.0&     255.620\\
        0.007&       -0.049&              82.3&     488.520\\
        0.007&       -0.049&              81.7&     688.760\\
        0.007&        0.101&              79.0&     130.260\\
        0.307&       -0.499&              76.4&     170.050\\
        0.307&       -0.499&              75.8&     206.370\\
        0.307&       -0.499&              75.1&     350.030\\
        0.307&       -0.499&              74.4&     200.560\\
        0.307&       -0.499&              73.8&     179.130\\
        0.307&       -0.499&              73.1&     141.600\\
     -15.309&     -12.949&            61.9&   154.000\\
     -15.309&     -12.949&            61.3&   108.320\\
\multicolumn{4}{l}{\bf G23.207$-$00.377}\\                                        
       -0.152&       -0.200&              92.9&      61.702\\
       -0.152&       -0.200&              92.2&     173.440\\
       -0.152&       -0.200&              91.6&     398.600\\
       -0.152&       -0.200&              90.9&     357.690\\
       -0.152&       -0.200&              90.3&     109.160\\
       -0.152&       -0.050&              81.0&     799.230\\
       -0.152&       -0.050&              80.4&    3030.800\\
       -0.152&       -0.050&              79.7&    6738.300\\
       -0.152&       -0.050&              79.1&    6281.900\\
       -0.002&       -0.050&              78.4&    3056.700\\
       -0.152&       -0.050&              77.8&    5250.100\\
       -0.002&       -0.050&              77.1&    6920.800\\
       -0.152&       -0.050&              76.4&    3010.400\\
      -1.505&       -0.950&               75.8&     172.700\\
       -0.152&       -0.050&              75.8&     217.300\\
       -0.152&       -0.050&              75.1&    1007.800\\
      -1.355&       -0.950&               75.1&     126.090\\
       -0.152&       -0.050&              74.5&    5069.900\\
       -0.152&       -0.050&              73.8&   10909.000\\
       -0.152&       -0.050&              73.2&   11463.000\\
       -0.152&       -0.050&              72.5&    5021.200\\
       -0.152&       -0.100&              71.8&     518.040\\
       -0.152&       -0.050&              71.2&     867.870\\
       -0.152&       -0.050&              70.5&    2283.600\\
       -0.152&       -0.050&              69.9&    2145.600\\
       -0.152&       -0.050&              69.2&     725.110\\
\hline
\end{tabular}
\end{table}
\begin{table}
\centering
\addtocounter{table}{-1}
\caption{continued.}
\begin{tabular}{rccr}
\hline\hline       
\\
$\Delta$RA & $\Delta$Dec & V$_{\rm LSR}$ & S \\
(\arcsec)     &  (\arcsec)      & (km\,s$^{-1}$)& (mJy\,beam$^{-1}$)\\
\hline
\multicolumn{4}{l}{\bf G23.207$-$00.377 {\it cont.}}\\
      -1.505&       -0.950&             69.2&     203.670\\
       -0.152&       -0.100&            68.5&     136.810\\
      -1.204&        0.400&             68.5&     153.120\\
      -1.505&       -0.950&             68.5&      79.018\\
      -1.204&        0.400&             67.9&      93.406\\
      -1.204&        0.400&             67.2&      60.951\\
       -0.152&       -0.050&            66.6&      62.789\\
      -1.204&        0.400&             65.9&      93.942\\
       -0.152&       -0.050&            65.9&      82.876\\
      -1.204&        0.400&             65.3&     238.600\\
      -1.204&        0.400&             64.6&     298.590\\
      -1.204&        0.400&             63.9&     164.130\\
\multicolumn{4}{l}{\bf G23.389$+$00.185}\\                                        
        0.003&       -0.050&            78.7&     127.570\\
       -0.147&       -0.050&            78.0&     145.710\\
       -0.147&       -0.050&            76.7&     141.910\\
        0.003&       -0.050&            76.1&      92.012\\
\multicolumn{4}{l}{\bf G23.707$-$00.198}\\
        0.901&        0.250&        74.6&     155.380\\
        0.751&        0.250&        73.9&      97.242\\
       3.753&      -4.850&          73.9&      67.946\\
       3.753&      -4.850&          72.6&     914.740\\
       3.753&      -4.850&          72.0&     682.860\\
        0.901&        0.250&           62.1&     131.370\\
        0.901&        0.250&           61.4&     278.160\\
        0.751&        0.250&           60.8&      89.936\\
\multicolumn{4}{l}{\bf G23.966$-$00.109}\\
       1.205&      -5.450&        77.2&     163.090\\
       1.055&        0.250&        73.9&     214.070\\
       1.055&        0.250&        73.3&     352.120\\
        0.005&       -0.050&         72.6&     774.970\\
        0.005&       -0.050&         72.0&     385.730\\
        0.005&       -0.050&         71.3&     177.550\\
        0.005&       -0.050&         70.7&     119.670\\
        0.005&       -0.050&         70.0&     143.980\\
        0.155&       -0.050&         69.3&     111.070\\
       1.055&        0.250&        54.9&     221.190\\
       1.055&        0.250&        54.2&     193.970\\
       1.055&        0.250&        53.5&     106.160\\
       1.055&        0.250&        52.9&     129.560\\
       1.055&        0.250&        51.6&     104.080\\
        0.905&        0.400&         50.9&     122.790\\
       1.055&        0.250&        48.9&    1120.800\\
       1.055&        0.250&        48.3&     255.340\\
       1.055&        0.250&        47.0&     209.790\\
       1.055&        0.250&        46.3&     110.470\\
       1.055&        0.250&        45.6&      73.665\\
        0.905&        0.400&         44.3&      91.386\\
       1.055&        0.250&        43.7&     138.280\\
        0.905&        0.250&         43.0&     130.530\\
       1.055&        0.250&        42.4&     106.440\\
       1.055&        0.250&        39.7&     111.540\\
       1.055&        0.250&        39.1&     180.410\\
       -0.295&       1.150&        36.4&     514.030\\
       -0.295&       1.150&        35.8&     851.210\\
       -0.295&       1.150&        35.1&     407.550\\
       -0.295&       1.150&        34.5&     233.570\\
\multicolumn{4}{l}{\bf G24.148$-$00.009}\\                                          
      14.281&      21.100&            24.4&     166.490\\
      14.281&      21.100&            23.7&      94.306\\
      14.432&      21.100&              0.03&      75.532\\
\hline
\end{tabular}
\end{table}
\begin{table}
\centering
\addtocounter{table}{-1}
\caption{continued.}
\begin{tabular}{rccr}
\hline\hline       
\\
$\Delta$RA & $\Delta$Dec  &V$_{\rm LSR}$ & S \\
(\arcsec)     &  (\arcsec)      &(km\,s$^{-1}$)& (mJy\,beam$^{-1}$)\\
\hline
\multicolumn{4}{l}{\bf G24.541$+$00.312}\\                                         
     -33.616&     -12.350&           101.1&     188.720\\
     -33.766&     -12.200&           100.4&     166.820\\
     -33.766&     -12.200&            99.8&      71.283\\
     -33.616&     -12.350&            99.1&     126.840\\
\multicolumn{4}{l}{\bf G26.598$-$00.024}\\                                          
        0.455&       -0.050&            22.2&    1088.200\\
       1.056&       -0.200&             29.5&      75.790\\
\multicolumn{4}{l}{\bf G28.817$+$00.365}\\                                          
       -0.615&       -0.449&       123.6&      73.343\\
       -0.615&       -0.449&       123.0&      74.491\\
       -0.615&       -0.049&        90.7&     232.440\\
       -0.615&       -0.049&        90.0&     606.260\\
       -0.615&       -0.049&        89.4&    1469.100\\
       -0.615&       -0.049&        88.7&    7338.900\\
       -0.615&       -0.049&        88.1&   13070.000\\
       -0.615&       -0.049&        87.4&    8506.800\\
       -0.615&       -0.049&        86.8&    2023.100\\
       -0.465&       -0.499&        86.1&     893.090\\
       -1.665&       -0.199&        86.1&     354.630\\
       -0.315&       -0.349&        85.4&    1251.800\\
       -1.665&       -0.199&        85.4&     285.680\\
       -0.465&       -0.499&        84.8&    1088.800\\
       -0.465&       -0.499&        84.1&     955.240\\
       -0.465&       -0.499&        83.5&     763.940\\
       -1.665&       -0.199&        83.5&     387.630\\
       -0.615&       -0.199&        82.8&     335.700\\
      -1.665&       -0.199&       82.8&     407.980\\
      -1.665&       -0.199&       82.1&     328.380\\
      -1.665&       -0.199&       82.8&     407.980\\
       -0.615&       -0.199&        82.8&     335.700\\
      -1.665&       -0.199&       82.1&     328.380\\
       -0.765&       -0.499&        81.5&     282.630\\
      -1.665&       -0.199&       81.5&     231.990\\
       -0.765&       -0.499&        80.8&     641.910\\
      -1.665&       -0.199&       80.8&     196.730\\
       -0.765&       -0.499&        80.2&     630.590\\
      -1.665&       -0.199&       80.2&     446.670\\
      -1.815&       -0.349&       79.5&     806.980\\
       -0.615&       -0.199&        79.5&     280.910\\
      -1.815&       -0.349&       78.9&     914.610\\
      -1.815&       -0.349&       78.2&     691.990\\
      -1.815&       -0.349&       77.5&     454.190\\
      -1.815&       -0.349&       76.9&     430.300\\
      -1.815&       -0.349&       76.2&     375.960\\
       -0.615&       -0.199&        76.2&     228.730\\
       -0.615&       -0.199&        75.6&     261.780\\
       -0.615&       -0.199&        74.9&     202.850\\
       -0.765&       -0.349&        69.0&     140.150\\
       -0.765&       -0.349&        68.3&     218.480\\
       -0.765&       -0.349&        67.7&     180.120\\
       -0.765&       -0.349&        59.1&     209.740\\
       -0.765&       -0.349&        58.4&     510.720\\
       -0.765&       -0.349&        57.8&     558.730\\
       -0.765&       -0.349&        57.1&     282.790\\
\multicolumn{4}{l}{\bf G30.318$+$00.070}\\                                         
       -0.903&        0.400&              52.6&     474.730\\
       -0.903&        0.400&              51.9&     291.770\\
       -0.903&        0.400&              51.2&     384.570\\
       -0.903&        0.400&              50.6&     235.530\\
       -0.903&        0.400&              49.9&    2807.700\\
       -0.903&        0.400&              49.3&    2283.700\\
       -0.903&        0.400&              47.9&    2274.500\\
       -0.903&        0.400&              47.3&    2013.200\\
       -0.903&        0.400&              46.6&     224.420\\
       -0.903&        0.400&              46.0&     878.560\\
       -0.903&        0.400&              45.3&     925.040\\
       -0.003&       -0.200&              32.8&     511.840\\
       -0.003&       -0.200&              32.2&     554.200\\
\hline
\end{tabular}
\end{table}

\begin{table}
\centering
\addtocounter{table}{-1}
\caption{continued.}
\begin{tabular}{rccr}
\hline\hline       
\\
$\Delta$RA & $\Delta$Dec  &V$_{\rm LSR}$ & S \\
(\arcsec)     &  (\arcsec) &(km\,s$^{-1}$)& (mJy\,beam$^{-1}$)\\
\hline
\multicolumn{4}{l}{\bf G30.400$-$00.296}\\                                         
       9.441&       4.751&           132.1&      46.758\\
       9.441&       4.751&           126.1&      68.852\\
       9.441&       4.751&           125.5&      77.645\\
       9.441&       4.751&           124.8&      53.516\\
       9.441&       4.751&           124.2&      52.276\\
\multicolumn{4}{l}{\bf G31.047$+$00.356}\\                                          
      -4.502&       1.300&             97.8&     173.050\\
      -4.802&       1.300&             78.1&     629.490\\
      -4.802&       1.300&             77.4&     669.660\\
      -4.652&       1.300&             76.8&      66.542\\
      -4.652&       1.300&             76.1&      91.005\\
      -4.652&       1.300&             72.1&      79.039\\
      -4.652&       1.300&             71.5&     558.540\\
      -4.652&       1.300&             70.8&     140.980\\
\multicolumn{4}{l}{\bf G31.581$+$00.077}\\                                          
        0.149&       -0.050&          103.5&     239.060\\
        0.149&       -0.050&          102.8&     621.510\\
        0.149&       -0.050&          102.2&     711.120\\
        0.149&       -0.050&          101.5&     350.840\\
      -4.051&       2.800&         100.9&    1698.600\\
      -4.051&       2.800&         100.2&    9004.900\\
      -4.051&       2.800&          99.5&   12545.000\\
      -4.051&       2.800&          98.9&    6135.200\\
      -4.051&       2.800&          98.2&    1318.600\\
      -4.051&       2.800&          97.6&     827.400\\
      -3.301&       3.400&          96.9&     536.440\\
      -3.901&       2.950&          96.9&     486.450\\
      -3.301&       3.400&          96.3&     275.000\\
      -2.401&       3.400&          95.6&     951.870\\
      -2.401&       3.400&          94.9&    1602.100\\
      -2.401&       3.400&          94.3&     789.870\\
        0.149&       -0.050&           91.7&     135.530\\
        0.149&       -0.050&           91.0&     136.240\\
        0.149&       -0.050&           90.3&     124.820\\
        0.149&       -0.050&           89.7&     121.650\\
\multicolumn{4}{l}{\bf G32.992$+$00.034}\\                                          
       -0.013&      -1.100&        85.9&     172.460\\
       -0.013&      -1.100&        85.2&     366.940\\
       -0.163&      -1.100&        83.9&     162.740\\
     -16.963&      25.450&         83.9&     500.950\\
     -16.963&      25.450&         83.2&    1070.200\\
     -16.963&      25.450&         82.6&     164.750\\
     -16.963&      25.450&         81.9&     135.560\\
     -16.963&      25.450&         77.3&     337.560\\
     -16.963&      25.450&         76.7&    1214.400\\
     -16.963&      25.450&         76.0&    1606.000\\
     -16.963&      25.450&         75.3&     532.860\\
     -16.963&      25.300&         74.7&     329.590\\
     -16.963&      25.450&         70.1&     324.050\\
     -16.963&      25.450&         69.4&     621.300\\
     -16.963&      25.450&         68.8&    1049.600\\
     -16.963&      25.450&         68.1&     537.610\\
     -16.963&      25.450&         65.5&     146.450\\
\multicolumn{4}{l}{\bf G33.641$-$00.228}\\                                          
        0.000&        0.100&             85.1&     767.180\\
        0.000&        0.100&             84.5&     892.410\\
        0.000&        0.100&             83.8&     211.630\\
        0.000&        0.100&             67.4&     316.000\\
        0.000&        0.100&             66.7&     318.810\\
        0.000&        0.100&             60.1&     274.990\\
        0.000&        0.100&             59.5&     705.260\\
        0.000&       -0.200&             58.8&     471.010\\
        0.000&       -0.200&             58.1&    1393.500\\
        0.000&       -0.200&             57.5&    2696.100\\
        0.000&       -0.200&             56.8&    5117.600\\
        0.000&       -0.200&             56.2&    3317.100\\
        0.000&       -0.200&             55.5&    3500.900\\
        0.000&       -0.200&             54.9&    21800.300\\
        0.000&       -0.200&             54.2&     745.680\\
\hline
\end{tabular}
\end{table}

\begin{table}
\centering
\addtocounter{table}{-1}
\caption{continued.}
\begin{tabular}{rccr}
\hline\hline       
\\
$\Delta$RA & $\Delta$Dec & V$_{\rm LSR}$ & S \\
(\arcsec)     &  (\arcsec)      &(km\,s$^{-1}$)& (mJy\,beam$^{-1}$)\\
\hline
\multicolumn{4}{l}{\bf G33.980$-$00.019}\\                                         
       -0.005&        0.250&             64.8&      56.244\\
       -0.005&        0.250&             64.2&      90.924\\
       -0.005&        0.250&             62.2&     398.350\\
       -0.005&        0.250&             61.5&     181.850\\
\multicolumn{4}{l}{\bf G34.751$-$00.093}\\                                       
       -0.014&        0.100&             74.4&     293.360\\
       -0.014&        0.100&             73.8&     379.520\\
       -0.164&       -0.200&             53.4&     344.340\\
       -0.164&       -0.200&             52.7&     525.630\\
       -0.164&       -0.200&             52.0&     200.420\\
       -0.014&        0.100&             48.1&     333.880\\
       -0.014&        0.100&             47.4&     249.800\\
       -0.014&        0.100&             45.5&     306.860\\
       -0.014&        0.100&             44.8&     331.100\\
       -0.014&        0.100&             42.8&     102.260\\
       -0.014&        0.100&             42.2&     123.410\\
       -0.014&        0.100&             38.2&      89.800\\
       -0.014&        0.100&             37.6&      97.497\\
\multicolumn{4}{l}{\bf G35.793$-$00.175}\\                                          
        0.150&        0.490&        64.6&     170.640\\
        0.150&        0.490&        64.0&     641.760\\
        0.150&        0.490&        63.3&    1645.500\\
        0.150&        0.490&        62.7&    1090.200\\
       -0.150&        0.040&        62.0&     637.780\\
        0.000&        0.340&        61.4&     205.560\\
      -1.350&        .790&       60.7&      67.977\\
      -1.350&        .790&       59.4&     571.250\\
      -1.350&        .790&       58.7&    2388.400\\
      -1.350&        .790&       58.1&     729.820\\
      -1.350&        .640&       49.5&     181.000\\
\multicolumn{4}{l}{\bf G36.115$+$00.552}\\                                         
       -0.157&        0.100&              79.6&     606.160\\
       -0.157&        0.100&              78.9&    1175.100\\
       -0.157&        0.100&              78.3&    4908.200\\
       -0.157&        0.100&              77.6&    38430.600\\
       -0.157&        0.100&              76.9&    1576.100\\
       -0.157&        0.100&              76.3&     260.940\\
       -0.007&        0.100&              73.7&     185.700\\
        0.293&        0.400&              40.7&     351.430\\
        0.293&        0.400&              40.1&     379.110\\
\multicolumn{4}{l}{\bf G36.705$+$00.096}\\                                       
       -0.163&        0.100&              68.2&      54.737\\
       -0.163&        0.100&              67.6&     106.380\\
        0.137&       -0.050&              56.4&     548.860\\
        0.137&       -0.050&              55.7&     332.540\\
\multicolumn{4}{l}{\bf G37.030$-$00.039}\\                                         
       -0.155&       -0.150&        81.9&      74.981\\
       -0.005&        0.000&        81.2&     266.990\\
       -0.155&        0.000&        80.6&     194.320\\
       -0.155&        0.000&        79.9&      87.324\\
       -0.005&        0.000&        77.3&     122.670\\
       2.847&      33.450&       77.3&     122.910\\
       -0.005&        0.000&        76.6&     212.280\\
       3.147&      33.750&       76.6&     111.340\\
       2.997&      33.600&       76.0&     102.090\\
       2.997&      33.600&       75.3&      85.365\\
       -0.906&      37.500&      75.3&      79.832\\
      -1.056&      37.350&       74.7&     260.000\\
      -1.056&      37.350&       74.0&      96.094\\
       2.847&      33.450&       72.0&      46.063\\
       2.997&      33.600&       71.4&      88.134\\
\hline
\end{tabular}
\end{table}
\begin{table}
\centering
\addtocounter{table}{-1}
\caption{continued.}
\begin{tabular}{rccr}
\hline\hline       
\\
$\Delta$RA & $\Delta$Dec  &V$_{\rm LSR}$ & S \\
(\arcsec)     &  (\arcsec)   &(km\,s$^{-1}$)& (mJy\,beam$^{-1}$)\\
\hline
\multicolumn{4}{l}{\bf G37.598$+$00.425}\\                                        
       -0.011&        0.100&          117.4&     219.240\\
       -0.011&        0.100&          116.7&     242.660\\
       -0.011&        0.100&          116.1&     318.110\\
       -0.011&        0.100&          114.8&    1031.300\\
       -0.011&        0.100&          114.1&    1030.300\\
       -0.011&        0.100&          102.9&     498.890\\
       -0.011&        0.100&          102.3&    2173.000\\
       -0.011&        0.100&          101.6&    2605.600\\
       -0.011&        0.100&          100.9&    2909.000\\
       -0.011&        0.100&          100.3&    1711.500\\
       -0.011&       -0.050&           99.6&     295.040\\
       -0.911&       -0.950&           92.4&     106.880\\
        1.940&      -2.600&            88.4&     173.860\\
        2.090&      -2.600&            87.8&     372.810\\
       -0.011&       -0.050&           77.9&    1780.700\\
       -0.011&       -0.050&           77.2&     761.960\\
\multicolumn{4}{l}{\bf G38.038$-$00.300}\\                                         
      15.451&      -3.801&           76.1&      69.425\\
      15.451&      -4.401&           75.4&     108.680\\
      15.301&      -3.801&           62.3&     149.360\\
      15.301&      -3.801&           61.6&      79.108\\
       -0.908&      13.749&          58.3&     141.580\\
\multicolumn{4}{l}{\bf G38.203$-$00.067}\\                                         
       -0.005&       -0.051&              90.8&     390.370\\
       -0.005&       -0.051&              90.1&     628.690\\
        0.145&       -0.201&              89.5&     510.060\\
       -0.005&       -0.051&              88.2&     158.110\\
       -0.005&       -0.051&              87.5&     235.350\\
       -0.005&       -0.051&              86.8&     154.100\\
       -0.005&       -0.051&              80.9&     120.230\\
       -0.005&       -0.051&              80.3&     161.100\\
        0.145&       -0.201&              79.6&     331.320\\
        0.145&       -0.201&              78.9&     275.340\\
\multicolumn{4}{l}{\bf G39.100$+$00.491}\\                                         
        0.295&        0.400&          24.5&     255.050\\
        0.595&        0.250&          23.9&     637.950\\
       -0.005&        0.850&          23.2&     223.880\\
       -0.456&       1.150&         22.5&     289.050\\
\\                                                         
\hline\hline                                               
\end{tabular}                                              
\end{table}

\end{document}